\documentclass[preprint,12pt]{elsarticle}
\usepackage[a4paper, total={6.5in, 8in}]{geometry}

\usepackage{lineno}
\usepackage{soul}
\usepackage[utf8]{inputenc}
\usepackage{mathtools}
\usepackage{amsthm}
\usepackage{tikz}

\usepackage{graphicx}
\usepackage{booktabs}
\usepackage{subcaption}
\usepackage{overpic} 

\DeclareMathAccent{\svec}{\mathord}{letters}{126}

\newenvironment{lequation*}{%
    \begin{linenomath*}\begin{equation*}%
}{%
    \end{equation*}\end{linenomath*}%
}

\renewcommand{\st}[1]{}

\usepackage{amssymb}

\usepackage{amsmath}
\usepackage{accents}
\usepackage{graphicx}
\usepackage[colorinlistoftodos]{todonotes}
\usepackage[colorlinks=true, allcolors=black]{hyperref}
\usepackage[hypcap=false]
{caption} 
\usepackage{subcaption}

\usepackage{listings}
\usepackage{url}
\usepackage{courier}
\usepackage{wrapfig}
\usepackage{color,soul}
\definecolor{codegreen}{rgb}{0,0.6,0}
\usepackage{chngpage}
\usepackage{xcolor}
\usepackage{bm}
\usepackage{amsmath,amssymb}
\usepackage{verbatim}
\newlength{\mycolwidth}
\usepackage{array}
\newcolumntype{Z}{>{$}p{\mycolwidth}<{$}}
\usepackage{algorithmicx}
\usepackage{algorithm} 
\usepackage{refcount}
\usepackage{algpseudocode} 
\usepackage{placeins} 
\MakeRobust{\Call}
\usepackage{amsfonts}
\usepackage{graphicx}
\usepackage{epstopdf}
\usepackage{amsmath} 

\usepackage{lscape}
\usepackage{booktabs}

\usepackage{tikz,lipsum,lmodern}
\usepackage[most]{tcolorbox}

\usepackage{pifont}

\DeclareMathAccent{\svec}{\mathord}{letters}{126}
\newcommand\acclrvec[1]{\accentset{\,\leftrightarrow}{#1}}
\newcommand\stvec[1]{\mathbf #1}
		
\newcommand\ssvec[1]{\acclrvec{\stvec{#1}}}	

\newcommand\cssvec[1]{\acclrvec{\tilde{\stvec{#1}}}}

\usepackage{scalerel,stackengine}
\stackMath

\usepackage{multirow}

\newcounter{bla}

\begin{document}

\begin{frontmatter}

\title{Can Explicit Subgrid Models Enhance Implicit LES Simulations?\\
A \textcolor{black}{Very }High-Order Solver Perspective}

\journal{Journal of Computational Physics}

\author[1,2]{Gonzalo Rubio\corref{cor1}}
\ead{g.rubio@upm.es}
\cortext[cor1]{Corresponding author.}
\author[3]{Gerasimos Ntoukas}
\author[1,2]{Miguel Ch\'avez-M\'odena}
\author[1,2]{Oscar Mari\~no}
\author[4]{Bernat Font}
\author[5]{Oriol Lehmkuhl}
\author[1,2]{Eusebio Valero}
\author[1,2]{Esteban Ferrer}



\address[1]{ETSIAE-UPM - School of Aeronautics, Universidad Polit\'ecnica de Madrid, Madrid, Spain}
\address[2]{Center for Computational Simulation, Universidad Polit\'ecnica de Madrid, Madrid, Spain}
\address[3]{Simerics Inc, Bellevue, WA, United States}
\address[4]{Faculty of Mechanical Engineering, Technische Universiteit Delft, The Netherlands}
\address[5]{Barcelona Supercomputing Center, Barcelona, Spain}

\begin{abstract}

\textcolor{black}{High-order Discontinuous Galerkin (DG) methods offer excellent accuracy for turbulent flow simulations and are increasingly attractive on modern GPU-oriented architectures, where high polynomial orders can improve arithmetic intensity for sufficiently large workloads. However, very high-order under-resolved simulations remain sensitive to the balance between numerical and modeled dissipation. Here, we investigate how explicit Vreman subgrid-scale (SGS) modeling interacts with the dissipation  introduced by split-form stabilization and Riemann solvers in a DGSEM framework. Using the three-dimensional Taylor--Green vortex at \(Re = 1600\) and an inviscid case (\(Re \rightarrow \infty\)), we evaluate kinetic energy dissipation, spectral accuracy, and numerical stability across well-resolved, under-resolved viscous, and strongly under-resolved regimes, comparing lower- and very high-polynomial-order configurations at comparable degrees of freedom.}

\textcolor{black}{Our results show that the usefulness of explicit SGS modeling depends strongly on the resolution regime, polynomial order, and numerical dissipation already present in the scheme. For the well-resolved configurations considered here, adding a Vreman model does not improve accuracy because the wavenumber range where the model acts overlaps with the inherent numerical dissipation of the DG scheme. Using a similar number of degrees of freedom, we observe that lower-order simulations introduce stronger numerical damping near the smallest represented scales, whereas very high-order simulations retain more spectral content but are more prone to high-wavenumber energy accumulation when insufficient dissipation is present. In contrast, when the resolution is insufficient, a suitably weak SGS contribution can help control this high-wavenumber energy accumulation, whereas excessive SGS dissipation degrades intermediate scales.}

\textcolor{black}{These findings identify regimes in which explicit SGS modeling is beneficial, neutral, or detrimental, and provide practical guidance for choosing dissipation mechanisms in very high-order DG turbulence simulations, which are well suited to modern GPU architectures.}

\end{abstract}


\end{frontmatter}

\section{Introduction}

High-order Discontinuous Galerkin (DG) methods have emerged as a powerful tool for the numerical simulation of complex fluid flows due to their ability to achieve arbitrary polynomial accuracy within each mesh element 
\cite{NodalDG,Moura2,Sherwin99dispersionanalysis,Juan_vonNeumann}. Unlike low-order schemes, high-order DG methods concentrate numerical errors at high wavenumbers, resulting in low dispersive and dissipative errors for well-resolved flows. This property ensures highly accurate representations of energy-containing scales and is particularly advantageous in Direct Numerical Simulation (DNS) and Large Eddy Simulation (LES) of turbulent flows.
\textcolor{black}{This high-order regime is also relevant from a computational perspective. Modern GPU architectures tend to favor arithmetic-intensive formulations, and sufficiently large high-order DG simulations can benefit from the increased arithmetic intensity associated with larger polynomial orders~\cite{Fehn2018,Gasparino2024SOD2D,kurz2025galaexi}. In this context, polynomial degrees around \(P=6{-}8\) have been reported to provide a favorable accuracy-throughput compromise in GPU-oriented high-order solvers, motivating the \(P=7\) configurations considered here.}

However, the high fidelity of DG methods also introduces challenges in under-resolved computations, such as implicit Large Eddy Simulations (iLES). In these situations, the grid or polynomial resolution is insufficient to capture the smallest turbulent scales, leading to numerical under-resolution and aliasing \cite{canuto_book}. Without adequate dissipation, energy accumulates at unresolved scales, potentially destabilizing the simulation; hence, under-resolved turbulent flows computed with high-order DG solvers require carefully designed stabilizing mechanisms that act locally near the cutoff scales.

A variety of stabilization strategies have been proposed to mitigate aliasing and improve robustness in high-order DG schemes. These include split forms or skew-symmetric variants of the convective terms \cite{Zang:1991:RSF:127290.127293,Blaisdell:1996:EFN:240862.240863}, interior penalty fluxes \cite{FerrerJCP}, over-integration \cite{Kirby_1,Kirby_0,Mengaldo2015}, and explicit or implicit filtering \cite{Blackburn2003610}. In particular, energy- and entropy-stable schemes exploit Gauss--Lobatto quadrature and the Summation--By--Parts Simultaneous-Approximation-Term (SBP--SAT) property \cite{FISHER2013518,doi:10.1137/130932193,SCHWARZ2025106874,manzanero2020entropy,manzanero2020free} to mimic continuous conservation of energy or entropy at the discrete level. These schemes provide intrinsic robustness, allowing under-resolved simulations to remain stable without relying on excessive numerical dissipation. Recent developments have extended these principles to the Spalart–Allmaras RANS equations \cite{LODARES2022110998} and to multiphase flows with p-adaptation \cite{ntoukas2021free,NTOUKAS2022111093}, demonstrating their versatility across flow regimes.

\textcolor{black}{The pioneering work of Flad and Gassner~\cite{FLAD2017782} showed that kinetic-energy-preserving split-form DG schemes provide a robust framework for LES of under-resolved turbulent flows. Their study clarified the role of interface Riemann-solver dissipation in iLES and demonstrated that, while this dissipation can act as an effective high-wavenumber filter in sufficiently resolved cases, it may become inadequate or poorly balanced at coarser LES resolutions. They also assessed explicit SGS modeling in this framework, mainly using Smagorinsky-type closures.} Subsequently, Winters et al.~\cite{WINTERS20181} further highlighted important limitations of iLES approaches when using high-order DGSEM with split forms in under-resolved turbulence. They observed that, at very high-orders (eighth order and above), stable solutions exhibited unphysical features, which were attributed to an energy-conserving bias resulting from the sharper dissipative behavior in wavenumber space. This limitation of iLES approaches has recently renewed interest in the use of explicit subgrid-scale (SGS) modeling within high-order discretizations.
Early LES developments showed that appropriate SGS closures, such as the classical Smagorinsky model \cite{smagorinsky1963general,lilly1965computational}, are crucial to stabilizing strongly under-resolved turbulent simulations. Subsequent formulations, including the WALE \cite{nicoud1999subgrid} and Vreman \cite{vreman2004eddy} models, aimed to reduce excessive damping and recover the correct near-wall scaling of the eddy viscosity.
\textcolor{black}{Whether explicit SGS modeling is beneficial is intrinsically related to the dissipative characteristics of the numerical scheme.} High-order DG methods are inherently low-dissipative, so the inclusion of a SGS model must carefully balance stability and spectral fidelity. Excess model dissipation can attenuate large-scale dynamics, while its absence may lead to an accumulation of energy near the grid cutoff \cite{garmann2013comparative,li2016priori}. Duan and Wang \cite{duan2024calibrating} recently argued that, in wall-modeled LES (WMLES), high-order methods benefit from additional SGS dissipation. They recommend using the Vreman model with a modified filter length that effectively increases the model constant compared to finite-volume implementations. In contrast, for wall-resolved LES (WRLES), a purely implicit approach remains preferable. This conclusion aligns with previous applications of the Vreman model in high-order WMLES studies, which have demonstrated improved stability and accuracy in under-resolved regimes \cite{Chatterjee2017ABL,mukha2024wall}.
Kumar et al. \cite{kumar2023turbulence} conducted a systematic study of WRLES using high-order schemes and found that the reduced model constants yield improved agreement with the reference data. Comparative assessments of explicit and implicit SGS strategies in DG formulations have also been reported in \cite{reddy2022comparison,Ntoukas2025F1}, underscoring the sensitivity of high-order LES to both modeling and discretization choices.
More recently, data-driven approaches have explored adaptive and discretization-consistent closures. Reinforcement learning has been used to dynamically adjust the SGS constant and enforce consistency between modeled and numerical dissipation \cite{Beck2023RLClosure,Kurz2023DRLLES}. These advances reflect a broader trend toward hybrid, adaptive SGS formulations capable of reconciling physical and numerical dissipation across scales.

\textcolor{black}{High-order DG methods have long been used for scale-resolving simulations, and very
high polynomial orders have already been explored in canonical turbulence benchmarks.
However, modern GPU-oriented implementations make such polynomial orders increasingly
attractive in practical simulations. This shift gives renewed
practical relevance to a known modeling question: how the dissipation supplied by the
numerical discretization interacts with explicit SGS modeling.} 
\textcolor{black}{At moderate polynomial orders, the broader numerical dissipation of the DG scheme and
the Riemann solver can mask or dominate the effect of an explicit eddy-viscosity model.
At very high-order, the numerical dissipation is more localized near the cutoff, so the
balance between Riemann-solver dissipation, split-form stabilization, and SGS viscosity
can change.}
\textcolor{black}{We study this dissipation balance using the three-dimensional Taylor--Green vortex at \(Re=1600\) and in the inviscid limit. The test matrix compares lower- and very high-polynomial-order configurations at comparable degrees of freedom, with particular emphasis on \(P=7\), corresponding to eighth-order accuracy. For the explicit SGS closure, we use the Vreman model, which is a robust eddy-viscosity formulation widely used in LES and suitable for high-order discretizations.
Within this framework, we first assess the role of split-form stabilization and then vary the Riemann solver and Vreman model constant to determine when the explicit SGS contribution complements the numerical dissipation, when it is essentially redundant, and when it overdamps resolved scales. The contribution of this work is a controlled assessment of explicit SGS modeling in very high-order DGSEM configurations of practical relevance for modern GPU-oriented solvers.}\\

The remainder of the paper is organized as follows. Section~\ref{sec:numerical_methodology} details the numerical methodology, and Section~\ref{sec:test_case_definition} introduces the test cases. The numerical experiments and results are presented in Section~\ref{sec:numerical_experiments} and the conclusions are summarized in Section~\ref{sec:conclusions}.

\section{Numerical Methodology}
\label{sec:numerical_methodology}
\textcolor{black}{All simulations in this work are performed using the high-order spectral element solver \href{https://github.com/horses-framework/horses3d-gpu}{\texttt{HORSES3D}} \cite{ferrer2023high,horses3dgpu}, developed at ETSIAE–UPM. The code implements the Discontinuous Galerkin Spectral Element Method (DGSEM) formulation on curvilinear hexahedral meshes of arbitrary order and supports  energy- and entropy-stable split forms, multiple Riemann solvers, the BR1 viscous discretization, explicit SGS models, and explicit Runge–Kutta time integration. \texttt{HORSES3D} is fully parallelized for large-scale computations on CPU and GPU clusters. Almost ideal scaling has been obtained in 20,000 CPU processors and 2,048 GPUs.}

\subsection{Compressible DGSEM Formulation}

The compressible Navier–Stokes equations are solved in conservative form (see \ref{sec:cNS}) using the DGSEM, a nodal high-order variant of the Discontinuous Galerkin (DG) method \cite{2009:Kopriva, ferrer2023high}. The DGSEM combines high-order accuracy with geometric flexibility and parallel scalability, making it well suited for large-scale turbulence simulations.

Two standard choices exist for the nodal distribution in DGSEM: Legendre--Gauss nodes and Legendre--Gauss--Lobatto nodes (hereafter, Gauss and Gauss--Lobatto nodes, respectively). Gauss nodes offer improved accuracy, while Gauss--Lobatto nodes enable slightly larger stable time steps for explicit time integrators \cite{doi:10.1137/100807211}. In this work, we employ Gauss nodes only for the preliminary assessment in Sec.~\ref{sec:preliminary-Gauss}, and adopt Gauss--Lobatto nodes for all subsequent simulations.

To ensure robustness in under-resolved conditions, the convective fluxes are discretized using energy- or entropy-preserving split forms, while controlled numerical dissipation is introduced via Riemann solvers (e.g., Roe-type fluxes). These intrinsic stabilization mechanisms can be complemented by explicit subgrid-scale (SGS) models for LES. The interplay between these three sources of dissipation—split forms, Riemann solvers, and SGS models—is systematically assessed in this work.

Viscous terms are discretized using the Bassi–Rebay 1 (BR1) scheme, which is equivalent to the interior penalty formulation when Gauss--Lobatto points and hexahedral elements are used \cite{MANZANERO20181}. The BR1 approach is neutrally stable \cite{2017:Gassner}, introducing only the minimal dissipation required for stability, unlike alternative formulations that include additional damping \cite{Unified}. A detailed description of the DGSEM formulation is provided in \ref{sec:DGSEM}.

\subsection{Energy- and Entropy-Stable Split Forms}\label{sec:stable}

High-order DG schemes can be formulated to preserve discrete analogues of continuous energy or entropy conservation by employing Gauss--Lobatto quadrature points for numerical integration. This property is enforced through the Summation--By--Parts Simultaneous Approximation Term (SBP--SAT) framework \cite{FISHER2013518,doi:10.1137/130932193,gassne_filter,Kopriva2,Juan_vonNeumann,10.1016/j.jcp.2016.09.013}, which enhances robustness in under-resolved turbulent flows and reduces aliasing errors \cite{DBLP:journals/jscic/ManzaneroRFVK18}.
Recent work has shown that the diagonal-norm SBP operator on Gauss nodes can also be expressed in flux-differencing form \cite{mateo2023flux}, thereby enabling the extension of stabilization techniques originally developed for Gauss--Lobatto nodes to Gauss nodes. Nevertheless, in this study we restrict ourselves to the traditional split-form formulations based on Gauss--Lobatto nodes. For an in-depth discussion of the respective advantages and drawbacks of split forms on Gauss and Gauss--Lobatto nodes, the reader is referred to the recent analysis in \cite{SCHWARZ2025106874}.

Extensive literature exists on energy- and entropy-stable DG schemes  \cite{FISHER2013518,doi:10.1137/130928650,10.1016/j.jcp.2016.09.013,10.1007/s10915-018-0702-1,MANZANERO2020109241,CHEN2017427,WINTERS20181,doi:10.1137/120890144}, with reviews available in \cite{10.1016/j.jcp.2016.09.013,winters2021construction,CSIAM-AM-1-1}. In these schemes, convective terms are rewritten in split form using two-point fluxes, which improves stability while preserving energy or entropy at the discrete level, at a modest additional computational cost. In this study, we compare several representative formulations, including the standard \cite{10.1016/j.jcp.2016.09.013}, Morinishi \cite{morinishi2010skew,10.1016/j.jcp.2016.09.013}, Ducros \cite{DUCROS2000114,10.1016/j.jcp.2016.09.013}, Kennedy–Gruber \cite{kennedy2008reduced}, Pirozzoli \cite{pirozzoli2010generalized}, entropy-conserving \cite{ismail2009affordable}, and Chandrasekar forms \cite{chandrashekar2013kinetic}.


\subsection{Riemann Solvers}

Riemann solvers are the standard approach to introduce numerical dissipation in DG schemes \cite{WorkshopDG,Gassner_iles}, as they arise naturally from the discretization of the non-linear convective terms. These solvers provide controlled dissipation, which can be further supplemented by turbulence models for high-Reynolds-number flows. Comparative studies of Riemann fluxes in homogeneous turbulence can be found in \cite{FLAD2017782,manzanero2018}.

The solvers evaluated in this study include both central and upwind-biased fluxes, covering a wide range of dissipative behaviors. The central flux represents the non-dissipative limit, while classical upwind formulations such as Roe’s approximate Riemann solver \cite{toro2013riemann} introduce controlled dissipation through characteristic decomposition. The low-dissipation Roe variant of Osswald et al.~\cite{osswald2016l2roe} and the matrix dissipation flux of Ismail and Roe \cite{ismail2009affordable} reduce numerical damping while retaining robustness in under-resolved regions. In contrast, the Lax–Friedrichs scheme \cite{Friedrichs1971} provides strong dissipation and is often used for stabilization in highly convective regimes.



\subsection{Subgrid-Scale Modeling}

For the explicit LES cases, we employ the Vreman SGS model \cite{vreman2004eddy}, selected for its robustness and proven suitability within high-order DG frameworks \cite{duan2024calibrating}. The model defines the eddy viscosity in a way that satisfies the physical requirement of vanishing turbulent viscosity near solid walls.
 
The SGS viscosity is expressed as
\begin{equation}
\begin{split}
\mu_t &= C_v \rho \sqrt{\frac{B_\beta}{\alpha_{ij}\alpha_{ij}}}, \quad
\alpha_{ij} = \frac{\partial u_j}{\partial x_i}, \\
\beta_{ij} &= \Delta^2 \alpha_{mi}\alpha_{mj}, \quad
\Delta = \frac{V^{1/3}}{P+1}, \\
B_\beta &= \beta_{11}\beta_{22}-\beta_{12}^2 + \beta_{11}\beta_{33}-\beta_{13}^2 + \beta_{22}\beta_{33}-\beta_{23}^2,
\end{split}
\end{equation}
where \( C_v \) is the model constant, \( V \) the element volume, and \( P \) the polynomial order. The influence of \( C_v \) on the solution is analyzed in this work, and the coupling of the SGS model to the compressible Navier–Stokes equations is detailed in \ref{sec:cNS}.

\subsection{Time Integration}

Temporal integration is performed using the low-storage, third-order explicit Runge–Kutta scheme (RK3) \cite{williamson1980low}. A fixed time step is employed, determined based on Courant-Friedrichs-Lewy (CFL) considerations to ensure stability. This scheme provides a balance between computational efficiency and temporal accuracy and is used consistently in all simulations presented.

\FloatBarrier
\section{Test Case Definition}
\label{sec:test_case_definition}

\subsection{Taylor--Green Vortex}

The canonical three-dimensional Taylor--Green vortex (TGV) at Reynolds numbers \( Re = 1600 \) and \( Re \rightarrow \infty \) is used to evaluate the performance of numerical schemes and their ability to capture the laminar–turbulent transition. In particular, the inviscid case is representative of viscous TGV solutions at very high Reynolds numbers \cite{WINTERS20181}.
\textcolor{black}{The domain is a cube of size \(L=2\pi\), discretized with a uniform Cartesian grid and periodic boundary conditions imposed on all sides. In Sec.~\ref{sec:preliminary}, different mesh resolutions and two representative polynomial orders, \(P=3\) and \(P=7\), are considered to assess their influence on accuracy and computational cost. After this preliminary analysis, the configuration is fixed to \(16^3\) hexahedral elements with polynomial order \(P=7\) within each element. The time step is selected according to the CFL stability restriction, with \(\Delta t = 6\times10^{-4}\) for the \(P3\) simulations and \(\Delta t = 2\times10^{-4}\) for the \(P7\) simulations. In addition, a time-step sensitivity study is reported in \ref{sec:app_dt}.}

\paragraph{\bf{Initial condition}}
The initial velocity and thermodynamic fields are set according to the standard divergence-free TGV formulation:

\begin{equation}
\begin{aligned}
\rho &= \rho_0, \\
u &= u_0 \, \sin\left(\frac{x}{L_0}\right) \, \cos\left(\frac{y}{L_0}\right) \, \cos\left(\frac{z}{L_0}\right), \\
v &= -u_0 \, \cos\left(\frac{x}{L_0}\right) \, \sin\left(\frac{y}{L_0}\right) \, \cos\left(\frac{z}{L_0}\right), \\
w &= 0, \\
p &= \frac{\rho_0 u_0^2}{\gamma M_0^2} + \frac{\rho_0 u_0^2}{16} \, 
\Big[ \big( \cos(2x/L_0) + \cos(2y/L_0) \big) \, \big( \cos(2z/L_0) + 2 \big) \Big],
\end{aligned}
\end{equation}

with $\gamma = 1.4$, $M_0 = 0.1$, $\rho_0 = 1$, $u_0 = 1$, and $L_0 = 1$. Time is non-dimensionalized as \(t/t_c\), where the characteristic time \(t_c = L_0 / u_0\).

The initial flow is smooth. As the simulation progresses beyond $t/t_c \gtrsim 6$, the flow transitions from the laminar initial state to a turbulent anisotropic regime, and beyond $t/t_c \gtrsim 13$, fully developed turbulence with isotropic structures emerges.

\paragraph{\bf{Diagnostics}}

To evaluate the performance of different numerical schemes, we monitor the \textit{kinetic energy} and its \textit{dissipation rate}, which provide insight into stability, energy transfer, and spectral fidelity. The kinetic energy is defined as
\[
E = \frac{1}{\rho_0 V} \int_V \frac{1}{2} \rho \, \mathbf{u} \cdot \mathbf{u} \, dV, \qquad 
\mathbf{u} = (u, v, w)^\mathrm{T}, \quad V = \text{domain volume},
\]
with dissipation rate
\[
\text{dissipation rate} = -\frac{dE}{dt}.\\
\]

Two complementary diagnostics are used to analyze each simulation:

\begin{itemize}
    \item \textbf{Kinetic energy dissipation rate over time} — indicates global energy decay and the stability of the numerical scheme.
    \item \textbf{Kinetic energy spectrum at $t/t_c = 9$} — quantifies the distribution of energy across scales, providing a measure of spectral fidelity.
\end{itemize}

The Nyquist wavenumber for the kinetic energy spectrum is
\begin{equation}
k_{\mathrm{Ny}} = \frac{\pi}{\Delta x}, \qquad 
\Delta x = \frac{L}{(P+1) N_{\mathrm{el}}},
\end{equation}
where \(L\) is the domain length, \(P\) the polynomial degree, and \(N_{\mathrm{el}}\) the number of elements per spatial direction \cite{fehn2022numerical,Bull2015TGV}. \textcolor{black}{Note that this Nyquist wavenumber is computed under the Fourier-basis assumption, corresponding to a two-points-per-wavelength (PPW) resolution limit. This estimate is optimistic for polynomial-based DG discretizations, since it overestimates the effective resolution capability of the polynomial approximation. The polynomial representation is limited by a smaller effective resolution limit, of around 3 PPW, so the usable spectral range should be interpreted as slightly lower than the nominal Nyquist value \cite{FLAD2017782}. For example, for the \(16^3\) elements, \(P=7\) simulations typically used in this work, the effective cutoff is approximately \(k_{\mathrm{eff}} \simeq 42.7\), rather than the nominal value \(k_{\mathrm{Ny}}=64\) shown in the spectra plots. Details on the computation of the spectra are provided in \ref{sec:app_fourier}.}

\paragraph{\bf{References}}

The kinetic energy dissipation of the TGV at $Re = 1600$ is compared against the dispersion-relation-preserving (DRP) solution on a $512^3$ mesh reported by Bull and Jameson~\cite{Bull2015TGV}.  
The energy spectra are compared with the pseudospectral results obtained on a $512^3$ grid by Carton de Wiart \textit{et al.}~\cite{CartonDeWiart2014DG}.  
For the inviscid TGV, the kinetic energy dissipation is compared with the reference data from Fehn \textit{et al.}~\cite{fehn2022numerical}, computed with an effective resolution of $8192^3$.  
The corresponding energy spectra are also compared with the same reference, using an effective resolution of $2048^3$.



\section{Numerical experiments}
\label{sec:numerical_experiments}

\textcolor{black}{This section investigates the dissipation mechanisms governing high-order DG large-eddy simulations of the three-dimensional Taylor--Green vortex. We begin with a preliminary assessment of the effects of the polynomial order, the SGS modeling, and the computational cost and compare discretizations with different combinations of mesh resolution and polynomial orders. This initial study highlights both the resolution advantages of very high-order discretizations (e.g., $P=7$) and the challenges associated with controlling dissipation in under-resolved simulations, thereby motivating the remainder of the work. It also assesses the CPU/GPU performance of the representative configurations, providing the practical motivation for focusing on very high-order discretizations in the remainder of the work.
Building on these observations, the subsequent sections systematically examine the individual and combined effects of split-form discretizations, Riemann solvers, and explicit SGS modeling on stability, energy dissipation, and spectral accuracy. Particular attention is paid to understanding how the dissipation introduced by each mechanism contributes to the overall behavior of the scheme and how these contributions balance one another in very high-order LES. Finally, the most promising configurations are assessed in the inviscid Taylor--Green vortex to further investigate the role of numerical dissipation in the absence of physical viscosity.}


\subsection{\textcolor{black}{Preliminary Assessment of Resolution, SGS Modeling, and Computational Cost}}
\label{sec:preliminary}


\subsubsection{\textcolor{black}{Impact of Spatial Resolution and Polynomial Order in TGV Simulations}}

In this section, we investigate the influence of mesh resolution and polynomial order in the Taylor--Green vortex at $Re=1600$. Two resolution levels are considered: a well-resolved configuration with approximately $2\times10^6$ DOF and an under-resolved configuration with approximately $2.5\times10^5$ DOF. The simulations are denoted as $N_{el}P$, where $N_{el}$ represents the number of elements per spatial direction and $P$ the polynomial order used in each element. Consequently, the well-resolved configurations are $32P3$ and $16P7$, while the under-resolved configurations are $16P3$ and $8P7$.
Unless otherwise stated, all simulations employ the Chandrasekar split-form DGSEM, the Roe Riemann solver, and, when applicable, the Vreman model with its standard coefficient, $C_v = 0.07$. Specific exceptions, including the use of the low-dissipation Roe (LD-Roe) flux and a reduced Vreman coefficient, $C_v = 0.01$, are explicitly identified where relevant. 

\paragraph{\textbf{ Well-resolved regime and effect of polynomial order}}

\begin{figure}[h!]
    \centering
    \begin{subfigure}[t]{0.49\textwidth}
        \includegraphics[width=\textwidth]{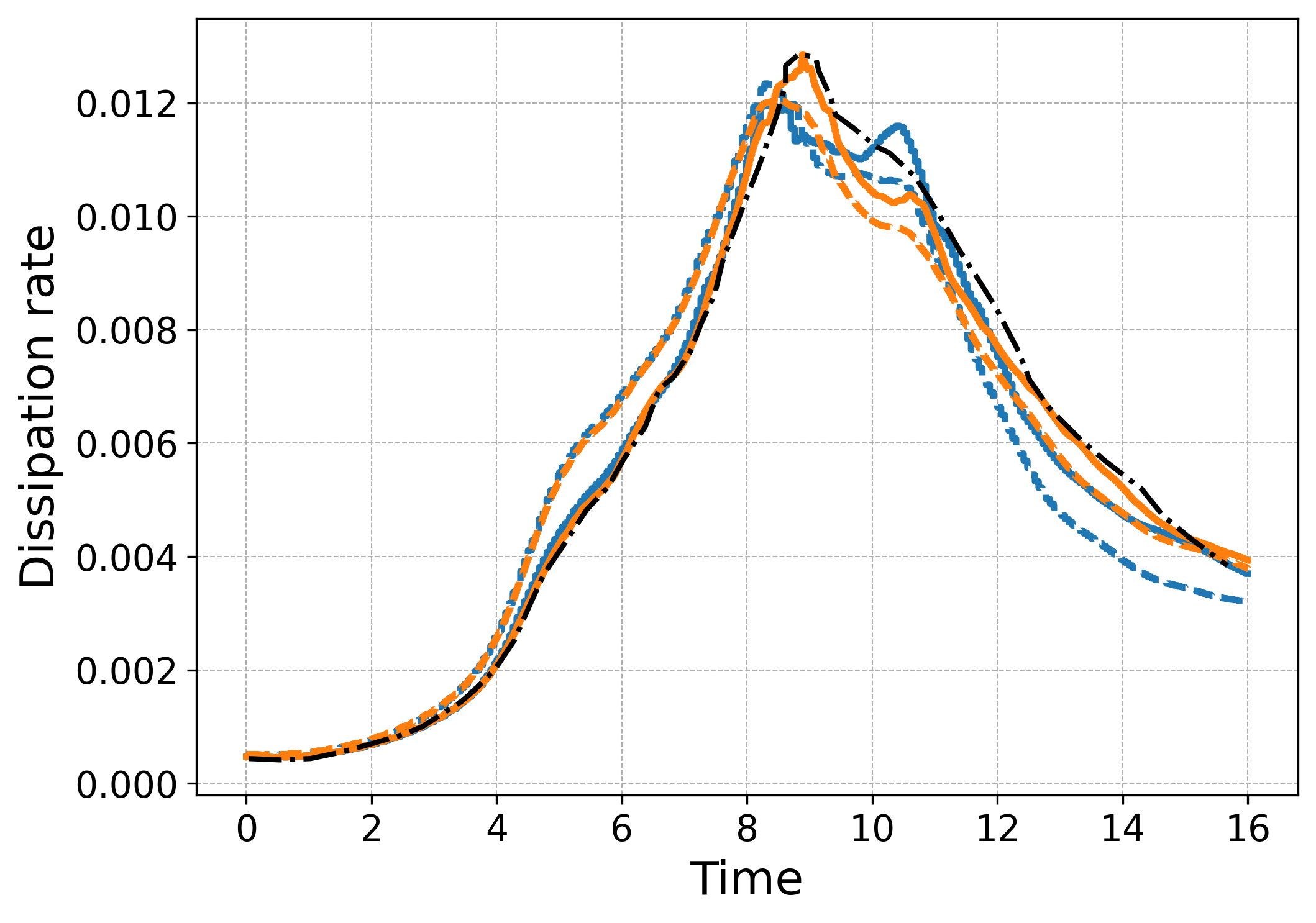}
        \caption{Kinetic energy dissipation rate.}
        \label{fig:32P3_vs_16P7_kinenrate}
    \end{subfigure}
    \hfill
    \begin{subfigure}[t]{0.49\textwidth}
        \includegraphics[width=\textwidth]{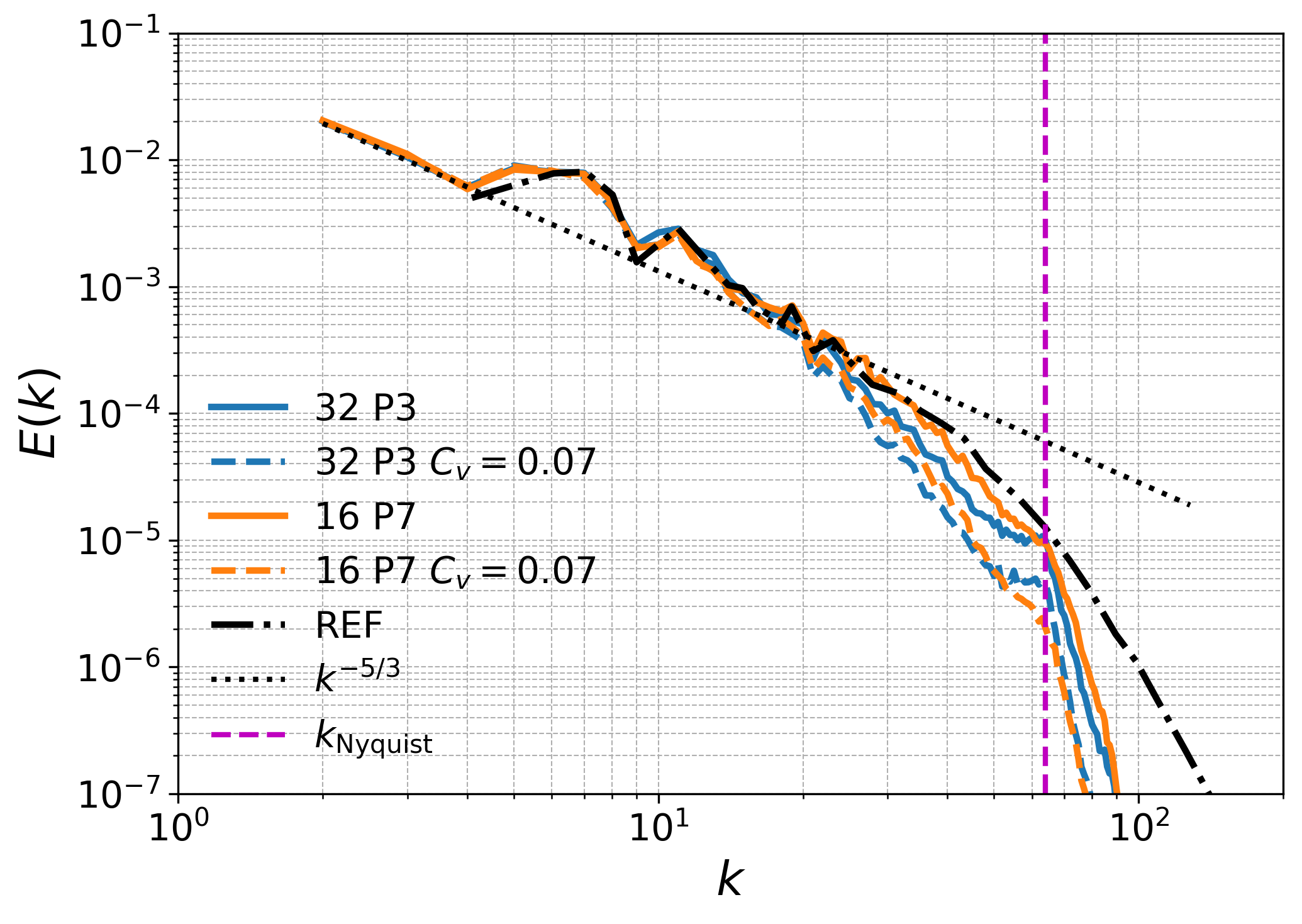}
        \caption{Kinetic energy spectrum at $t/t_c=9$.}
        \label{fig:32P3_vs_16P7_spectra}
    \end{subfigure}
    \caption{Effect of polynomial order in a well-resolved setting on dissipation and spectral behavior. Comparison of $32P3$ and $16P7$ configurations with and without a turbulence model. For the metrics considered here, the higher-order iLES configuration provides the closest agreement with the reference \cite{Bull2015TGV,CartonDeWiart2014DG}, while the standard Vreman coefficient introduces excessive damping.}

    \label{fig:comp1_32P3_vs_16P7}
\end{figure}

We first compare, in Fig.~\ref{fig:comp1_32P3_vs_16P7}, the $32P3$ and $16P7$ configurations in the iLES setting. The figure shows the time evolution of kinetic energy dissipation rate and the kinetic energy spectrum. For an equivalent number of degrees of freedom, the higher-order discretization provides a more accurate representation of the flow evolution. This improvement is evident in both the kinetic energy dissipation rate (Fig.~\ref{fig:32P3_vs_16P7_kinenrate}) and the energy spectra (Fig.~\ref{fig:32P3_vs_16P7_spectra}), where the $16P7$ simulation agrees more closely with the reference solution.
The energy spectra are presented in Fig.~\ref{fig:32P3_vs_16P7_spectra}. Both discretizations accurately reproduce the reference spectrum up to approximately $k \approx 20$, with $16P7$ providing the closest overall agreement. At higher wavenumbers, the differences between the two approaches become more apparent. The $32P3$ discretization exhibits a more dissipative behavior, consistent with the larger numerical dissipation associated with lower-order approximations and the Roe flux. In contrast, the $16P7$ simulation retains more energy and achieves improved spectral fidelity over a wider range of resolved scales.

However, the improved resolution capability of the $P7$ discretization is accompanied by a slight accumulation of energy near the cutoff wavenumbers. Although this effect is relatively small in the present case, it suggests that the numerical dissipation provided by the baseline discretization may be insufficient to fully control the highest resolved scales. This observation will become relevant in the subsequent analysis, where the interplay between numerical and modeled dissipation is investigated in detail.

When the Vreman model is activated using its standard coefficient, $C_v=0.07$, the accuracy of both discretizations deteriorates. The impact is particularly noticeable during the transitional phase, where additional SGS dissipation suppresses part of the resolved dynamics and leads to a poorer prediction of the dissipation peak. The effect is even more evident in the energy spectra, where the model removes energy over a broad range of resolved scales rather than acting primarily near the cutoff wavenumbers.

This behavior is especially detrimental for the $P7$ discretization. Because of its higher resolution capability, a larger fraction of the turbulent spectrum is explicitly resolved, making the solution more sensitive to excessive SGS dissipation. As a result, the standard Vreman coefficient degrades the spectral accuracy of the high-order simulations and largely removes the advantages provided by the increased polynomial degree. For the cases considered here, the iLES approach consistently yields the most accurate results.

\paragraph{\textbf{Under-resolved regime and effect of polynomial order}}

\begin{figure}[h!]
    \centering
    \begin{subfigure}[t]{0.49\textwidth}
        \includegraphics[width=\textwidth]{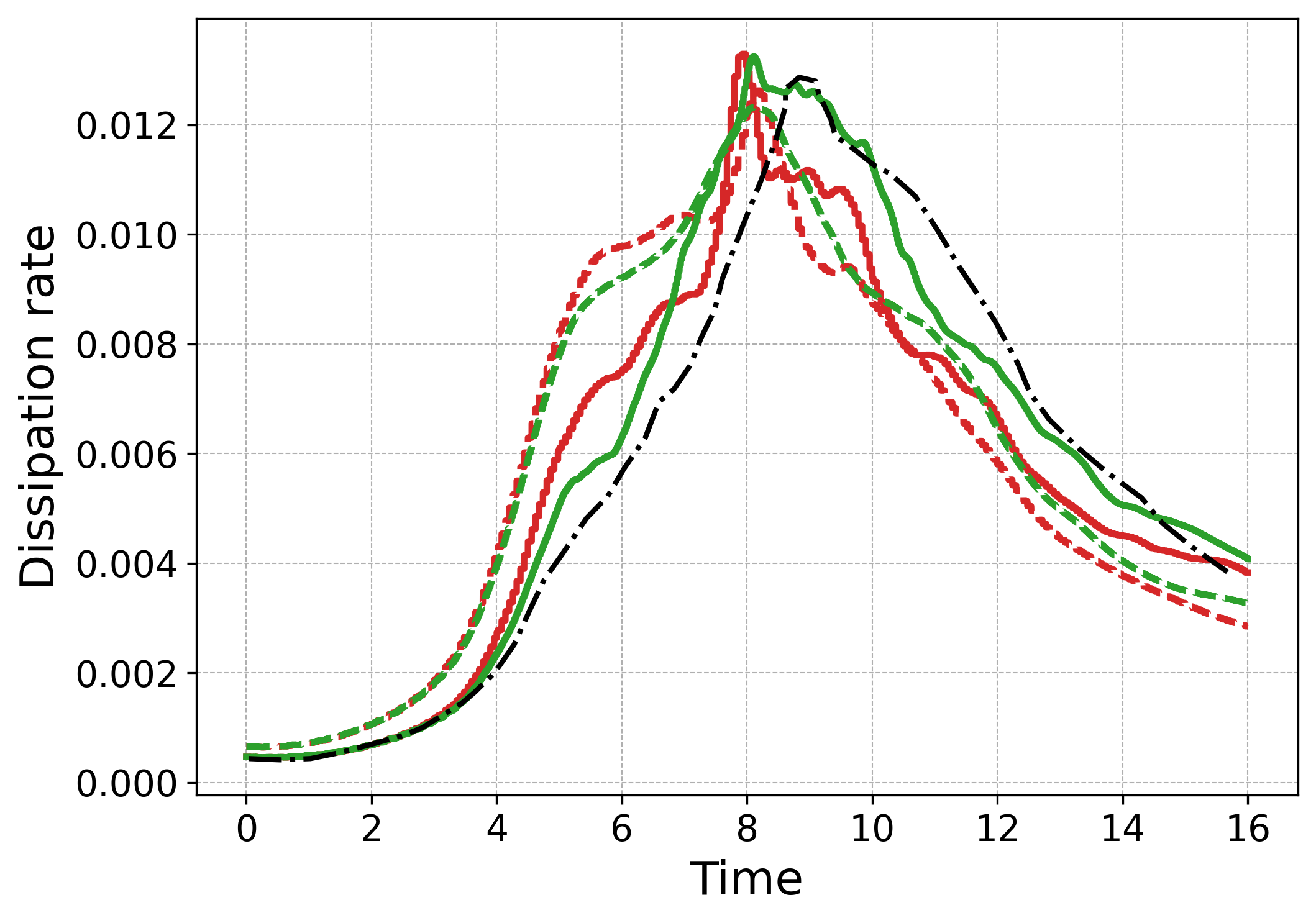}
        \caption{Kinetic energy dissipation rate.}
        \label{fig:comp2_16P3_vs_8P7_kinenrate}
    \end{subfigure}
    \hfill
    \begin{subfigure}[t]{0.49\textwidth}
        \includegraphics[width=\textwidth]{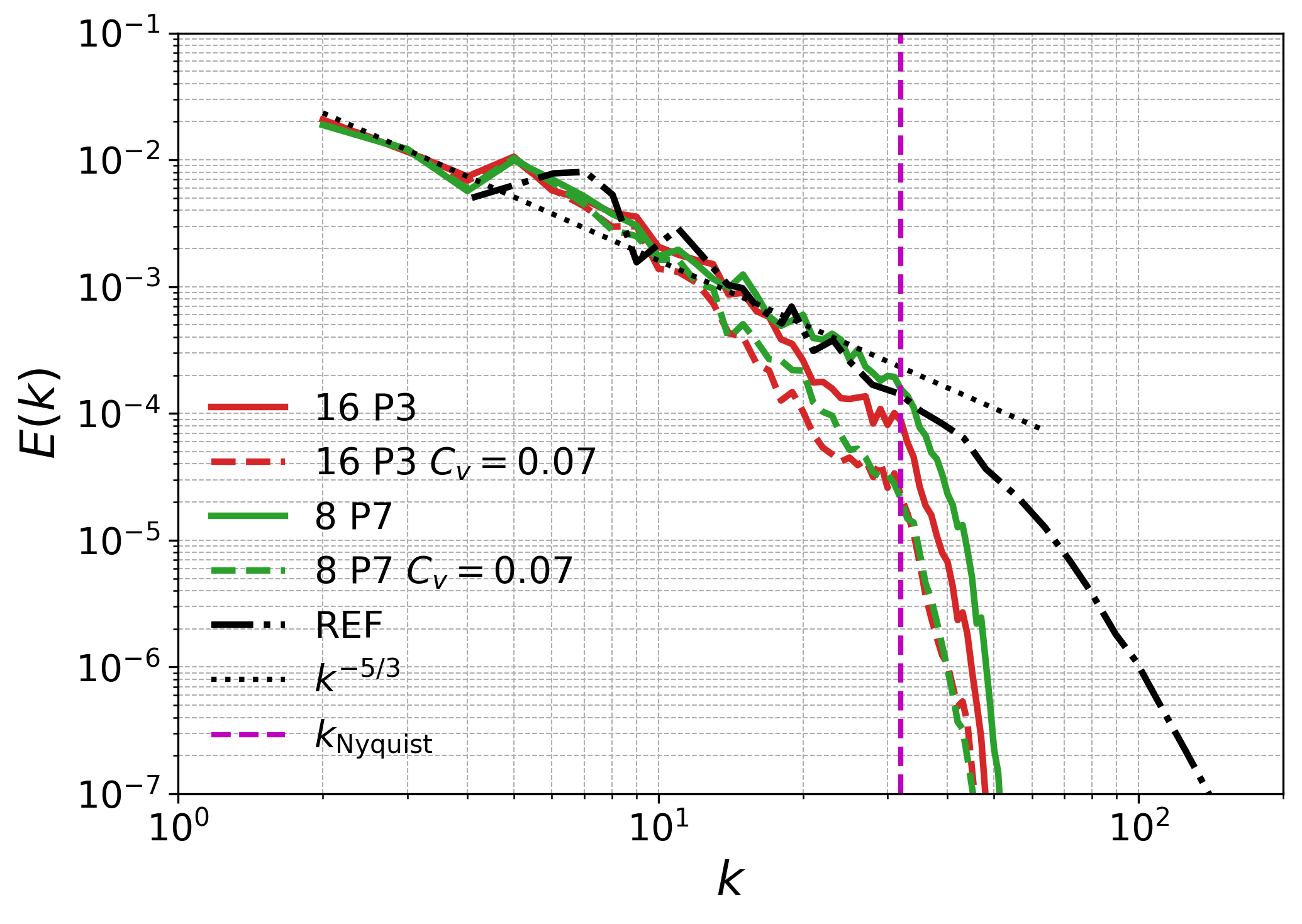}
        \caption{Kinetic energy spectrum at $t/t_c=9$.}
        \label{fig:comp2_16P3_vs_8P7_spectra}
    \end{subfigure}
    \caption{Effect of polynomial order in an under-resolved setting on dissipation and spectral behavior. Comparison of the $16P3$ and $8P7$ configurations with and without a turbulence model. The higher-order discretization retains more energy across the spectrum and provides improved spectral fidelity despite the reduced number of elements. However, a slight accumulation of energy near the cutoff wavenumbers is observed for the $P7$ configuration, while the standard Vreman coefficient introduces excessive dissipation and degrades the solution in both cases.}

    \label{fig:comp2_16P3_vs_8P7}
\end{figure}

Second, we compare the $16P3$ and $8P7$ configurations in the iLES setting. The evolution of the dissipation rate, shown in Fig.~\ref{fig:comp2_16P3_vs_8P7_kinenrate}, reveals significant deviations from the reference solution for both discretizations, indicating that neither resolution is sufficient to accurately capture the flow dynamics. Nevertheless, the $8P7$ configuration consistently remains closer to the reference than the $16P3$ configuration, suggesting that increasing the polynomial order provides a more effective use of the available degrees of freedom in this under-resolved regime.

The corresponding kinetic energy spectra are presented in Fig.~\ref{fig:comp2_16P3_vs_8P7_spectra}. In the iLES simulations, the $16P3$ simulation is systematically more dissipative than the $8P7$ one, especially across the high-wavenumber ranges. As a result, the $8P7$ configuration provides a more accurate representation of the energy distribution throughout the reliably resolved range. However, a slight accumulation of energy is observed near the Nyquist cutoff, indicating that the numerical dissipation provided by the split form and Riemann solver is insufficient to fully control the smallest resolved scales. As discussed later for the inviscid TGV case (see Sec.~\ref{sec:InviscidTGV}), this lack of dissipation becomes more pronounced and may lead to degraded solution quality, consistent with the observations reported in \cite{WINTERS20181}.

When the Vreman model is activated with its standard coefficient, the agreement with the reference solution deteriorates further, particularly during the laminar and transitional stages, as shown in Fig.~\ref{fig:comp2_16P3_vs_8P7_kinenrate}. The spectral results in Fig.~\ref{fig:comp2_16P3_vs_8P7_spectra} reveal that both discretizations exhibit a very similar response once the subgrid-scale model is introduced. In this regime, the additional dissipation imposed by the model dominates the numerical dissipation mechanisms, excessively damping the intermediate scales and suppressing the energy content over a substantial portion of the resolved spectrum. Again, the effect is less noticeable for $P=3$, as the baseline scheme was already more dissipative. 
These results indicate that, for such coarse discretizations, the standard value of the Vreman coefficient leads to an overly dissipative solution and masks the potential benefits associated with increasing the polynomial order.\\


\paragraph{\textbf{Dissipation balance in under-resolved high-order LES}}

The previous comparison indicates that the standard Vreman coefficient ($C_v=0.07$) introduces excessive dissipation in both the well-resolved and under-resolved simulations at $Re=1600$. A natural question is therefore whether a weaker SGS contribution can improve the under-resolved solution without degrading the resolved scales. To investigate this idea, we consider the most challenging viscous configuration studied in this section, namely the $8P7$ discretization, and compare Roe, LD-Roe, and their combinations with a weak Vreman model ($C_v=0.01$).

\begin{figure}[h!]
    \centering
    \begin{subfigure}[t]{0.49\textwidth}
        \includegraphics[width=\textwidth]{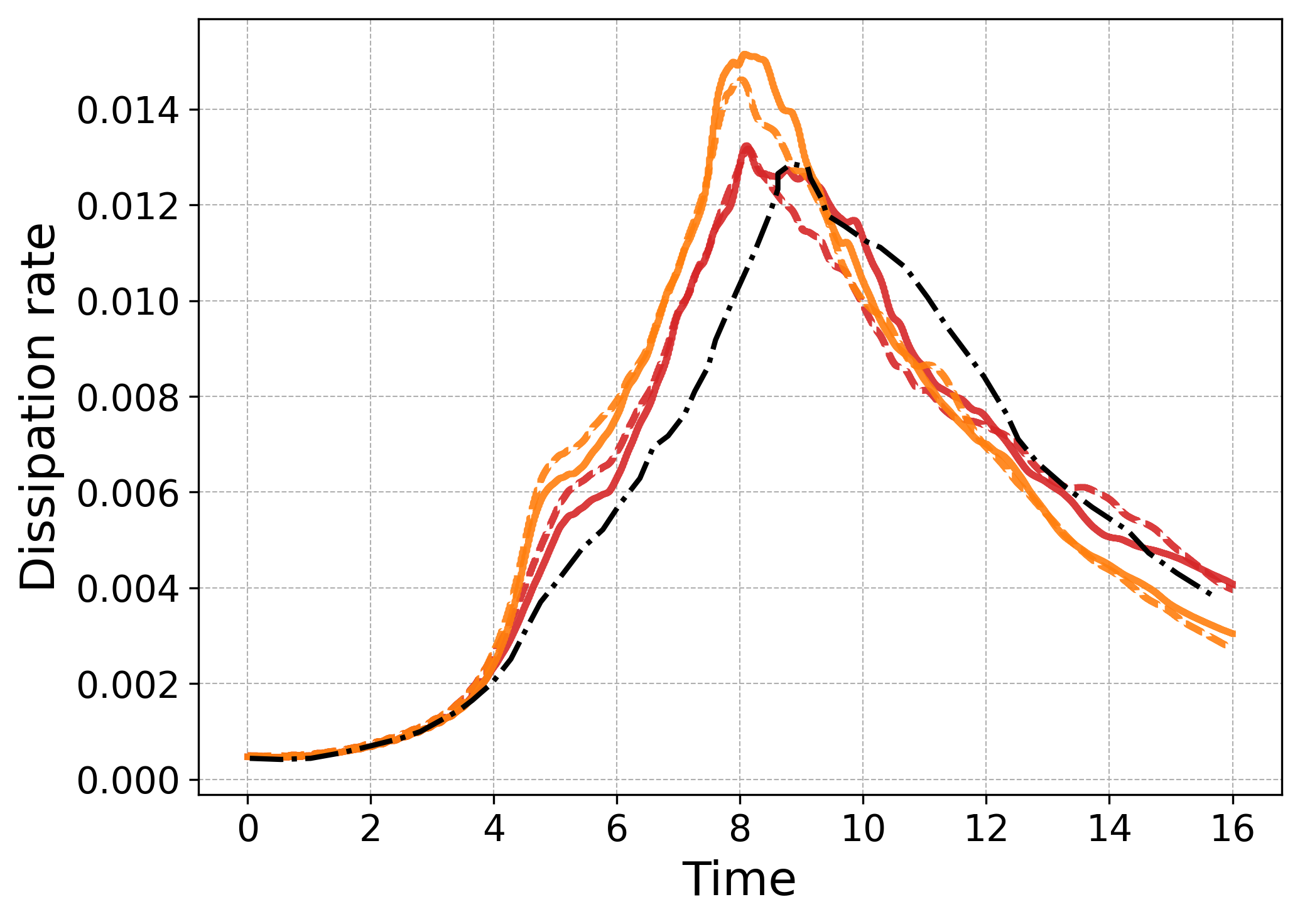}
        \caption{Kinetic energy dissipation rate.}
        \label{fig:underresolved_re1600_diss}
    \end{subfigure}
    \hfill
    \begin{subfigure}[t]{0.49\textwidth}
        \includegraphics[width=\textwidth]{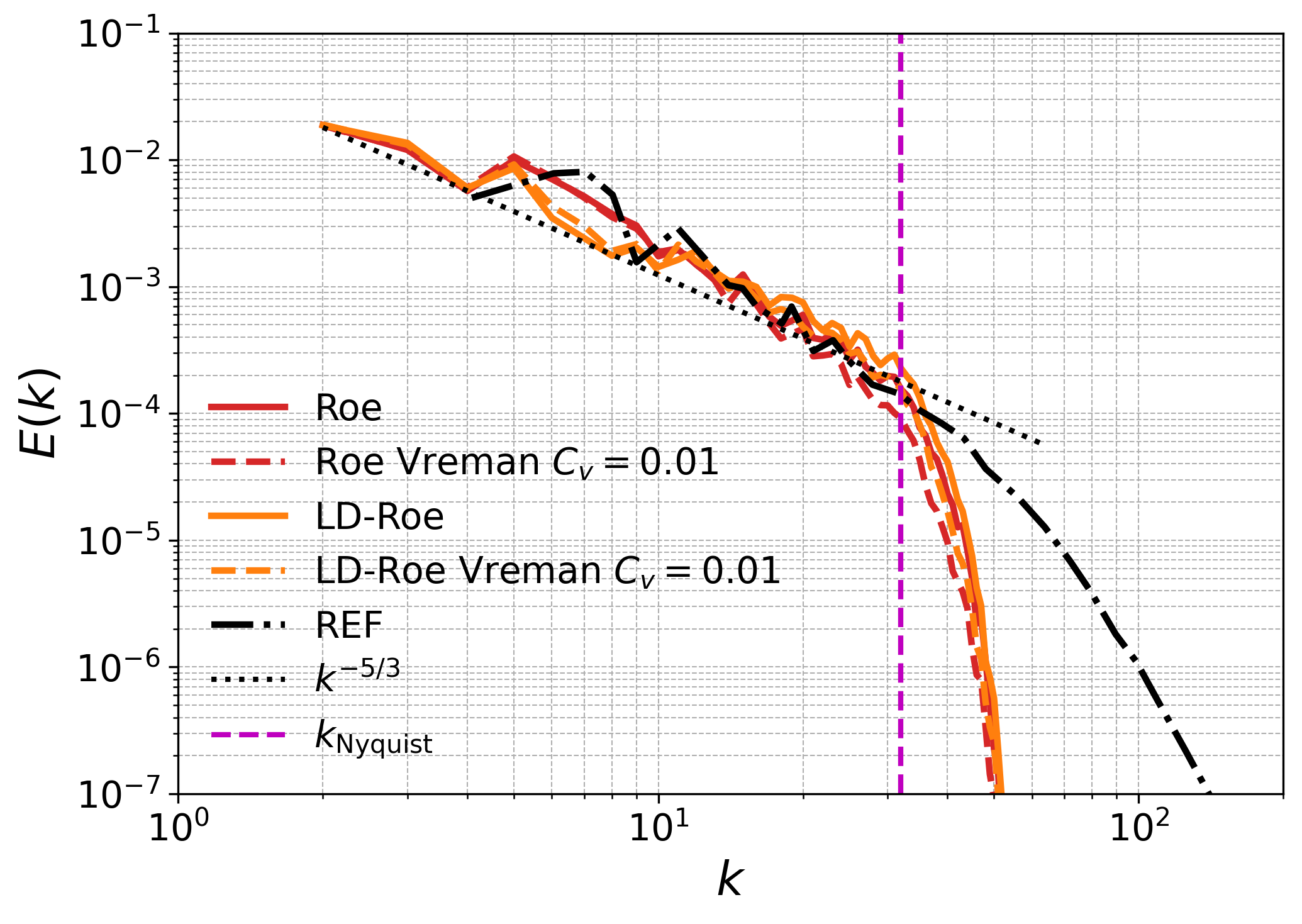}
        \caption{Kinetic energy spectrum at $t/t_c=9$.}
        \label{fig:underresolved_re1600_spec}
    \end{subfigure}
    \caption{Effect of weak SGS dissipation in an under-resolved $Re=1600$ simulation. Comparison of Roe and LD-Roe fluxes with and without the Vreman model ($C_v=0.01$) for the $8P7$ configuration. The weak SGS model has a limited impact on the global dissipation rate but improves the spectral behavior near the cutoff wavenumbers, particularly when combined with the low-dissipation Roe flux, resulting in a more balanced distribution of numerical and modeled dissipation.}

    \label{fig:underresolved_re1600}
\end{figure}

The results are shown in Fig.~\ref{fig:underresolved_re1600}. The kinetic energy dissipation rate, Fig.~\ref{fig:underresolved_re1600_diss}, indicates that all configurations reproduce the overall temporal evolution of the flow. However, differences become apparent during and after flow transition. The low-dissipation Roe flux has a higher dissipation rate than the standard Roe formulation, while the addition of the weak SGS model provides a moderate increase in dissipation.

The energy spectra at $t/t_c=9$, shown in Fig.~\ref{fig:underresolved_re1600_spec}, provide further insights. The Roe scheme is slightly more dissipative across the upper part of the spectrum, whereas LD-Roe retains additional energy close to the cutoff wavenumbers. When combined with the weak-constant SGS model, the excess high-wavenumber energy is reduced while preserving the improved spectral resolution of the low-dissipation flux. 

Although explicit SGS modeling is unnecessary for the well-resolved configurations considered in this work, a weak SGS contribution can be beneficial in under-resolved high-order simulations by providing controlled dissipation near the cutoff scales without the excessive damping associated with the standard model coefficient.


In summary, for a fixed number of degrees of freedom, high-order discretizations with fewer elements provide improved resolution capabilities. This advantage is most evident in the laminar and transitional regimes, where the $P7$ configurations considered here remain closer to the reference than their lower-order counterparts. 

Once turbulence is fully developed, the \(P7\) discretizations retain more spectral content over the reliably resolved range, but they can also exhibit energy accumulation near the cutoff. When this accumulation is controlled with the Vreman model, the choice of model constant becomes important. The standard Vreman coefficient introduces excessive damping, whereas the under-resolved \(8P7\) simulations indicate that a weaker SGS contribution can be beneficial when combined with numerical fluxes, providing additional control of the highest resolved wavenumbers without significantly compromising the resolved scales.


\subsubsection{\textcolor{black}{Computational cost on modern CPU/GPU architectures}}

\begin{table}[t]
\centering
\caption{Performance comparison between CPU and GPU executions on MareNostrum~5 for a representative set of DGSEM simulations. The table reports measured wall-clock time and the performance index (PID), see Eq.~\ref{eq:PID}. All simulations consist of 1000 time steps, and the reported values are averaged over three independent runs. The last column reports the PID ratio between CPU and GPU executions, interpreted as the effective CPU-core-equivalent acceleration factor provided by a single GPU.}
\label{tab:cpuvsgpuperf}
\begin{tabular}{lccrcccc}
\hline
Elem & $P$ & DoF 
& \multicolumn{2}{c}{Wall time (s)} 
& \multicolumn{2}{c}{PID (s)} 
& PID ratio \\
\cline{4-5}\cline{6-7}
 & & & CPU & GPU & CPU & GPU & CPU/GPU \\
\hline
$16^3$ & 3 & 262{,}144      
& $1.44 \times 10^{1}$  
& $4.51 \times 10^{0}$  
& $2.06 \times 10^{-6}$ 
& $2.30 \times 10^{-8}$ 
&  89.53 \\

$16^3$ & 7 & 2{,}097{,}152   
& $1.20 \times 10^{2}$  
& $1.56 \times 10^{1}$  
& $2.14 \times 10^{-6}$ 
& $9.90 \times 10^{-9}$ 
& 216.21 \\

$32^3$ & 3 & 2{,}097{,}152   
& $1.55 \times 10^{2}$  
& $2.38 \times 10^{1}$  
& $2.77 \times 10^{-6}$ 
& $1.51 \times 10^{-8}$ 
& 182.82 \\

$32^3$ & 7 & 16{,}777{,}216  
& $9.88 \times 10^{2}$  
& $9.81 \times 10^{1}$  
& $2.20 \times 10^{-6}$ 
& $7.80 \times 10^{-9}$ 
& 281.98 \\

$64^3$ & 3 & 16{,}777{,}216  
& $1.27 \times 10^{3}$  
& $1.75 \times 10^{2}$  
& $2.82 \times 10^{-6}$ 
& $1.39 \times 10^{-8}$ 
& 203.23 \\

$64^3{}^{\dagger}$ & 7 & 134{,}217{,}728 
& $1.96 \times 10^{3}$  
& $3.80 \times 10^{2}$  
& $2.18 \times 10^{-6}$ 
& $7.55 \times 10^{-9}$ 
& 288.53 \\
\hline
\multicolumn{8}{l}{\footnotesize \(^{\dagger}\) This case was run using 4 CPU nodes and 2 GPU nodes; the rest used 1 CPU node and 1 GPU node.} \\
\end{tabular}
\end{table}

We simulate the same cases on CPUs and GPUs to illustrate the practical performance motivation for considering very high-order discretizations on modern GPU architectures. For this comparison, the discretization is fixed to the Chandrasekar split form with the Roe flux in the iLES setting.


The results compare executions on MareNostrum~5. The CPU runs use general-purpose (GPP) nodes, each equipped with 2$\times$ Intel Sapphire Rapids 8480+ CPUs at 2~GHz, with 56 cores per socket and 112 cores per node. The GPU runs use accelerated (ACC) nodes equipped with 4 NVIDIA Hopper GPUs per node.

All cases in Table~\ref{tab:cpuvsgpuperf} are run using one CPU node and one ACC node, except for the largest case, \(64^3\) elements with \(P=7\), which is run using 4 CPU nodes and 2 GPU nodes due to memory requirements. 
We report wall-clock time (in seconds) together with the performance index (PID)~\cite{Krais2021FLEXI}, defined as the time required to advance one degree of freedom through a single Runge–Kutta stage on one processing unit (CPU core or GPU):
\begin{equation}
\label{eq:PID}
\text{PID} = \frac{\text{Walltime} \times \#\text{Ranks}}{\#\text{RK stages} \times \text{DoF} \times \#\Delta t},    
\end{equation}
where $\#\text{RK stages}=3$ for the RK3 time integration scheme. All simulations were run for $\#\Delta t = 1000$ time steps, and the reported values correspond to averages over three independent runs. The monitoring routines were disabled to avoid perturbing the performance measurements. The timings correspond to the full explicit time-integration procedure, including residual evaluations and Runge--Kutta stage updates, rather than to isolated kernel benchmarks.

In Table~\ref{tab:cpuvsgpuperf}, we report wall-clock times and performance indices for a representative set of simulations. The magnitude of the measured PID values is consistent with values reported for closely related DGSEM implementations~\cite{kurz2025galaexi}.
For a fixed number of degrees of freedom, both CPU and GPU results indicate improved computational efficiency for higher polynomial orders. Simulations with higher $P$ and fewer elements exhibit lower wall-clock times than configurations with lower $P$ and more elements. This behavior is attributed to the increased arithmetic intensity of high-order discretizations, which improves the arithmetic-to-memory operation ratio, increases vectorization efficiency, and enables larger and more efficient matrix–vector operations. 
An additional effect of the GPU-based configuration is the reduction in the total number of processes required to execute a given simulation, which reduces MPI communication overheads and improves scalability under GPU-to-GPU data transfer (avoiding passing through CPU host).

When comparing GPU and CPU performance, a clear dependence on polynomial order is observed. For $P=7$, each GPU provides a performance equivalent to approximately 280--290 CPU cores. Accordingly, a single GPU node (4 GPUs) is comparable to around ten CPU nodes (1,120 cores). A reduction in performance is observed for the \(16^3\) elements, \(P=7\) case. This is because near-optimal GPU utilization in the present implementation requires a minimum workload of approximately $10,000$ elements per GPU.

For $P=3$, the corresponding ratio decreases to approximately 180--200 CPU cores per GPU. In this regime, a single GPU node is comparable to about seven CPU nodes (784 cores). For the smallest case ($16^3$, $P=3$), the ratio further decreases to approximately 90 CPU cores per GPU, consistent with insufficient workload per device. Overall, within the present implementation and benchmark, high-order simulations provide significant computational advantages on GPUs and motivate the use of the $P=7$ configurations examined below.

In terms of energy consumption, and taking into account the measured node power under the High-Performance Linpack (HPL) benchmark reported in \cite{Banchelli2025MN5}, simulations on GPU nodes are also more efficient. On MareNostrum~5, one ACC node consumes approximately 3.5~kW under HPL, whereas one GPP node consumes approximately 0.95~kW with both sockets fully loaded. The PID already encodes the resource-time (ranks × walltime) per degree of freedom and Runge--Kutta stage, so the CPU/GPU PID ratio directly measures the CPU-core-equivalents delivered by one GPU. Assuming the minimum workload of 10,000 elements per GPU, for $P=7$, this ratio is $\sim280$, so a single four-GPU node is equivalent to $280\times4=1,120$ CPU cores, i.e., around ten fully loaded GPP nodes ($1,120/112=10$). In the conservative limit where the GPUs operate at the thermal design power (maximum capacity), and noting that the power ratio between CPU and GPU nodes is 0.95~kW / 3.5~kW $=$ 0.27, $P=7$ GPU simulations are $10\times0.27\approx2.7$ times more energy-efficient than CPU ones. For $P=3$, this factor drops to 1.93.


\subsubsection{\textcolor{black}{Discussion and motivation}}

The results presented in this section provide several observations that motivate the remainder of this work. First, for a fixed number of degrees of freedom, increasing the polynomial order generally improves the representation of both the transient flow dynamics and the turbulent energy spectrum. This behavior is observed in both the well-resolved and under-resolved configurations considered here and is consistent with the enhanced resolution properties commonly associated with high-order DG discretizations.

At the same time, increasing the polynomial order alone is not sufficient to guarantee improved predictions. While the \(P=7\) discretizations provide a more accurate representation of the resolved scales, the under-resolved simulations exhibit signs of energy accumulation near the cutoff range when only numerical dissipation is present. This behavior is less pronounced for \(P=3\), for which the larger amount of numerical dissipation at small wavelengths provides additional damping near the cutoff range. This suggests that very high-order discretizations may place increased demands on the mechanisms responsible for controlling the smallest represented scales.

A second observation concerns the role of SGS modeling. For the well-resolved configurations considered in this study, the iLES approach yields the closest agreement with the reference solution, indicating that additional SGS dissipation is not required at these resolutions. In contrast, the under-resolved cases suggest that some form of SGS dissipation may be beneficial. However, the results also show that the standard Vreman coefficient, \(C_v=0.07\), introduces excessive damping for the discretizations examined here, leading to a degradation of both the dissipation-rate evolution and the energy spectra. These findings indicate that the effectiveness of SGS modeling depends strongly on its interaction with the dissipation already introduced by the numerical scheme.

Finally, the performance analysis provides an additional practical motivation for studying these effects. The high-order discretizations considered here achieve both improved spectral resolution and increased computational efficiency on modern GPU architectures. The role of the GPU results is therefore to explain why very high-order configurations are attractive in production settings, not to suggest that GPUs create new turbulence-modeling physics. As such architectures continue to play an increasingly important role in large-scale CFD simulations, understanding how to combine very high-order discretizations with appropriate dissipation mechanisms becomes increasingly relevant.

Taken together, these observations suggest that the dissipation balance in very high-order LES deserves further investigation. In the present framework, dissipation is introduced through the split-form discretization, the Riemann flux, and the SGS model. The relative contribution of each mechanism, and the extent to which they can be combined to provide sufficient robustness without unnecessarily degrading the resolved scales, remains unclear. The following sections therefore examine these components individually and in combination in order to better understand their role in high-order LES.

\subsection{Well-Resolved LES (The Viscous Taylor–Green Vortex): Implicit and Explicit Dissipation Mechanisms}

We consider the Taylor–Green Vortex at Reynolds 1600, representing a well-resolved LES. \textcolor{black}{In this section and throughout the remainder of the paper, we retain the \(16^3\), \(P=7\) configuration identified in the preliminary assessment. }
The goal is to assess whether iLES can be improved by adding an explicit SGS model. Stability is first ensured using split forms, after which we compare dissipation mechanisms for accuracy, contrasting implicit iLES (Riemann fluxes) with explicit SGS contributions. Finally, a hybrid approach combining Riemann solvers and SGS modeling is tested to evaluate potential improvements.

\subsubsection{Standard vs Split-Form Discretizations}
\label{sec:preliminary-Gauss}
In this preliminary analysis, we evaluate the effect of split-form (e.g., Chandrasekar)  formulations in comparison to standard discretizations. Note that the split-form schemes employ Gauss--Lobatto nodes, while in this subsection the standard discretization uses Gauss nodes. 

Simulations using the standard approach exhibit numerical instabilities and eventually diverge. In particular, the central configuration becomes unstable around $t/t_c \approx 4$, while the Roe variant remains stable for a longer period but fails near $t/t_c \approx 7$. These observations indicate that, in the absence of split-form stabilization, aliasing errors accumulate as nonlinear interactions intensify, ultimately leading to blow-up during the transitional regime. The Riemann solver alone does not provide sufficient dissipation to stabilize the simulation.

\begin{figure}[h!]
    \centering
    \includegraphics[width=0.49\textwidth]{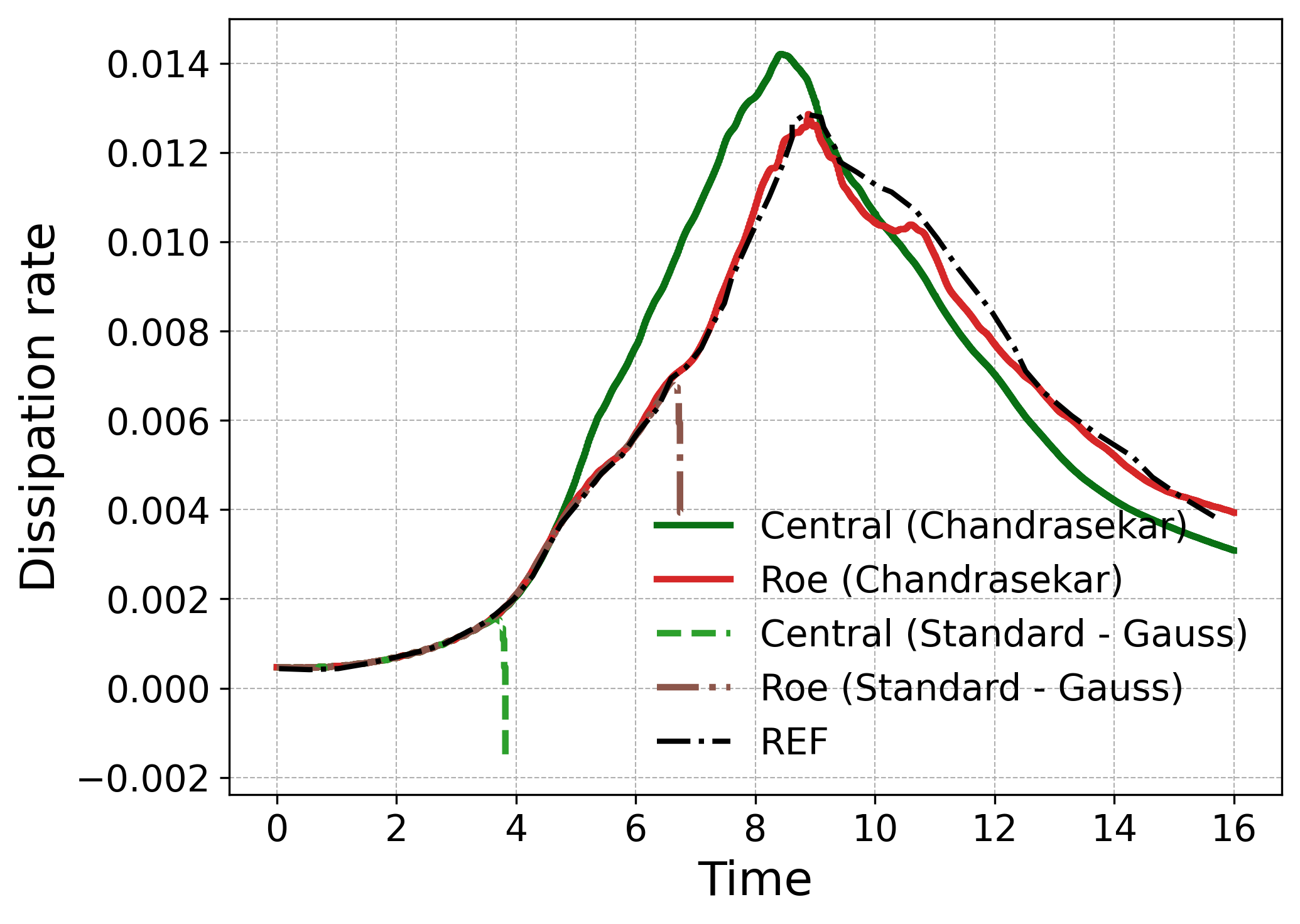}
    \caption{Kinetic energy dissipation rate for standard (non-split) formulations with Gauss nodes. Both central and Roe-based fluxes eventually diverge due to aliasing-driven instabilities, while the split-form scheme introduces stabilizing dissipation. \textcolor{black}{The Roe--Chandrasekar split-form configuration remains stable and closely follows the reference solution over the full time interval.}}
    \label{fig:part0_kinenrate}
\end{figure}

Fig.~\ref{fig:part0_kinenrate} presents the time evolution of the kinetic energy dissipation rate, highlighting the onset of instability for both standard discretizations. For comparison, the figure also includes the behavior of a representative split-form scheme \textcolor{black}{combined with central and Roe Riemann solvers}. Around $t/t_c \approx 5$, the split form introduces additional dissipation, effectively stabilizing the simulation and preventing blow-up. \textcolor{black}{The combination of the split form with the Roe Riemann solver follows the reference closely while remaining stable over the full time interval, showing that the combination of split-form stabilization and Roe interface dissipation provides a robust baseline for the subsequent analysis.}

These findings highlight the critical role of split-form discretizations in ensuring robustness for under-resolved simulations. 
By introducing additional nonlinear stabilization, split forms effectively control aliasing errors throughout the laminar–turbulent transition.

\subsubsection{Performance of Central Split Forms}

This part investigates the performance of multiple central split-form discretizations without explicit SGS modeling. As in the remainder of the paper, all formulations---including the standard (non split-form) discretization---employ Gauss--Lobatto nodes. Interface fluxes are computed using minimally dissipative central fluxes combined with the BR1 scheme for viscous terms. This setup isolates the intrinsic stabilization provided by each split form.

\begin{figure}[h!]
    \centering
    \begin{subfigure}[t]{0.49\textwidth}
        \includegraphics[width=\textwidth]{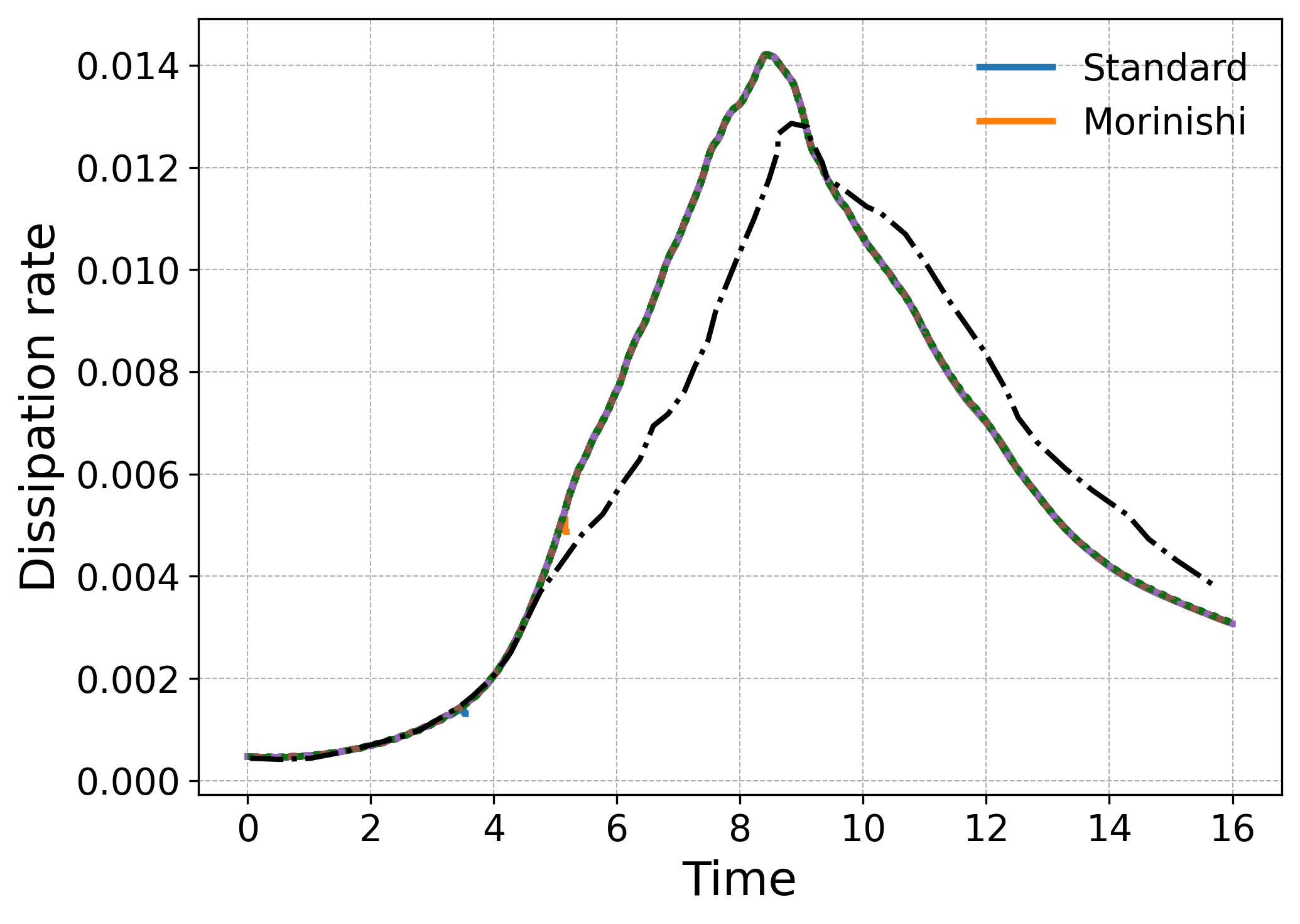}
        \caption{Kinetic energy dissipation rate.}
        \label{fig:part1_kinenrate}
    \end{subfigure}
    \hfill
    \begin{subfigure}[t]{0.49\textwidth}
        \includegraphics[width=\textwidth]{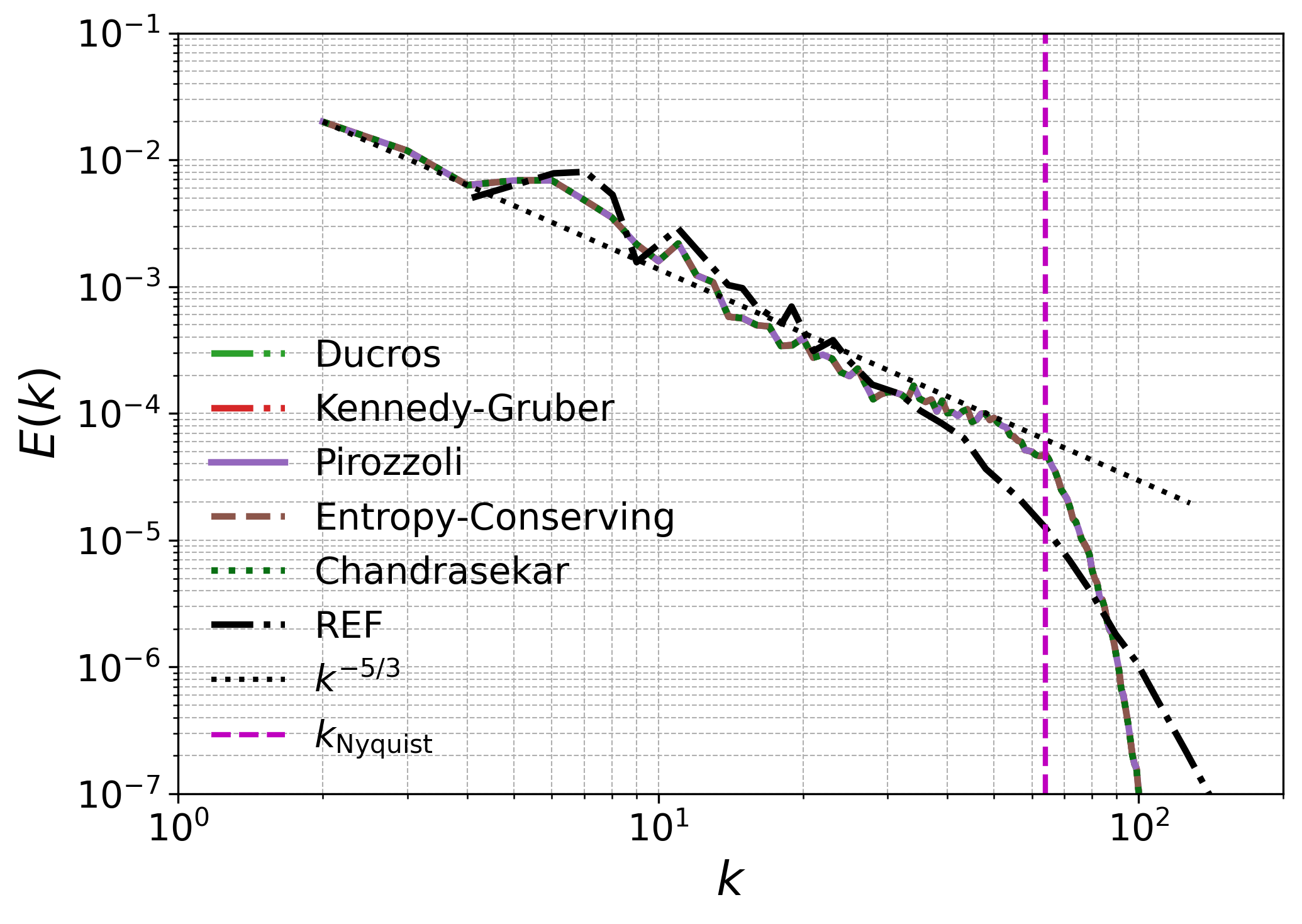}
        \caption{Kinetic energy spectrum at $t/t_c=9$.}
        \label{fig:part1_spectra}
    \end{subfigure}
    \caption{Effect of central split forms on dissipation and spectral behavior. The standard and Morinishi schemes are shown only in Fig.~\ref{fig:part1_kinenrate}, as they became unstable before $t/t_c=9$. Results for all schemes are shown in both figures, and the curves overlap, indicating only minimal differences between the methods. Overall, central split forms exhibit excess dissipation during transition and insufficient dissipation once turbulence is fully developed.}
    \label{fig:part1}
\end{figure}

Among the tested formulations, Ducros, Kennedy--Gruber, Pirozzoli, Entropy-conserving, and Chandrasekar split forms remain stable throughout the simulation and exhibit nearly identical kinetic energy dissipation rates (Fig.~\ref{fig:part1_kinenrate}). In contrast, the Standard and Morinishi schemes become unstable early, at $t/t_c \approx 3.5$ and $t/t_c \approx 5$, respectively.

During the transitional phase ($t/t_c \in [4,8]$), all stable split forms display a dissipation peak exceeding that in the reference data. This enhanced dissipation is attributable to aliasing-driven implicit numerical dissipation, which compensates for the under-resolution of small scales, as noted in \cite{Bull2015TGV}.

In the fully turbulent regime ($t/t_c > 8$), the kinetic energy decays smoothly and the resulting spectra closely follow the reference slopes for moderate wavenumbers. At high wavenumbers ($k > 30$), some energy accumulation is observed due to insufficient dissipation in the smallest resolved scales (Fig.~\ref{fig:part1_spectra}).

Overall, central split forms provide intrinsic stabilization sufficient for under-resolved simulations. However, when used without additional dissipation mechanisms, they tend to over-dissipate energy during the transitional phase and under-dissipate at high wavenumbers in the turbulent regime, affecting the spectral distribution of energy.
Most split-form discretizations exhibited comparable behavior for this low-Mach-number case, with the exception of the standard and Morinishi formulations.
For the remainder of the paper, the Chandrasekar split form is adopted, as it has previously been shown to provide improved robustness in complex and highly compressible flows \cite{mateo2025unsupervised,SCHWARZ2025106874}.

\subsubsection{Impact of the Vreman SGS Model}

In this section, the Chandrasekar split form is combined with an explicit Vreman SGS model of varying strength to evaluate its influence on the accuracy of the numerical scheme. The Vreman model introduces tunable dissipation at unresolved scales~\cite{vreman2004eddy} (see \ref{sec:cNS} for details). Two model constants are considered: $C_v = 0.07$, the standard value in finite-volume methods~\cite{vazquez2016alya}, and $C_v = 0.01$, a smaller value suggested for high-order methods~\cite{kumar2023turbulence}.

\begin{figure}[h!]
    \centering
    \begin{subfigure}[t]{0.49\textwidth}
        \includegraphics[width=\textwidth]{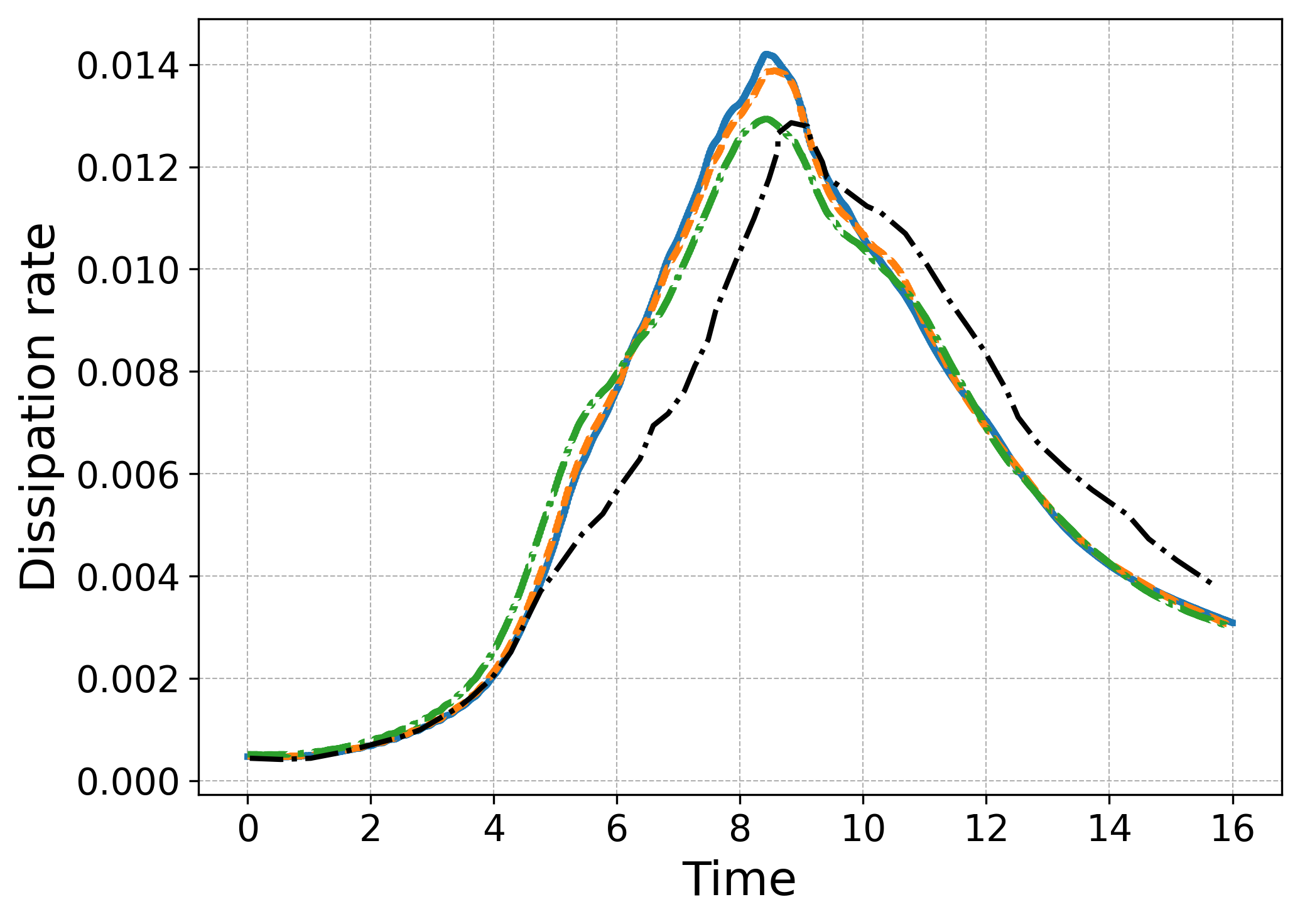}
        \caption{Kinetic energy dissipation rate.}
        \label{fig:part2_kinenrate}
    \end{subfigure}
    \hfill
    \begin{subfigure}[t]{0.49\textwidth}
        \includegraphics[width=\textwidth]{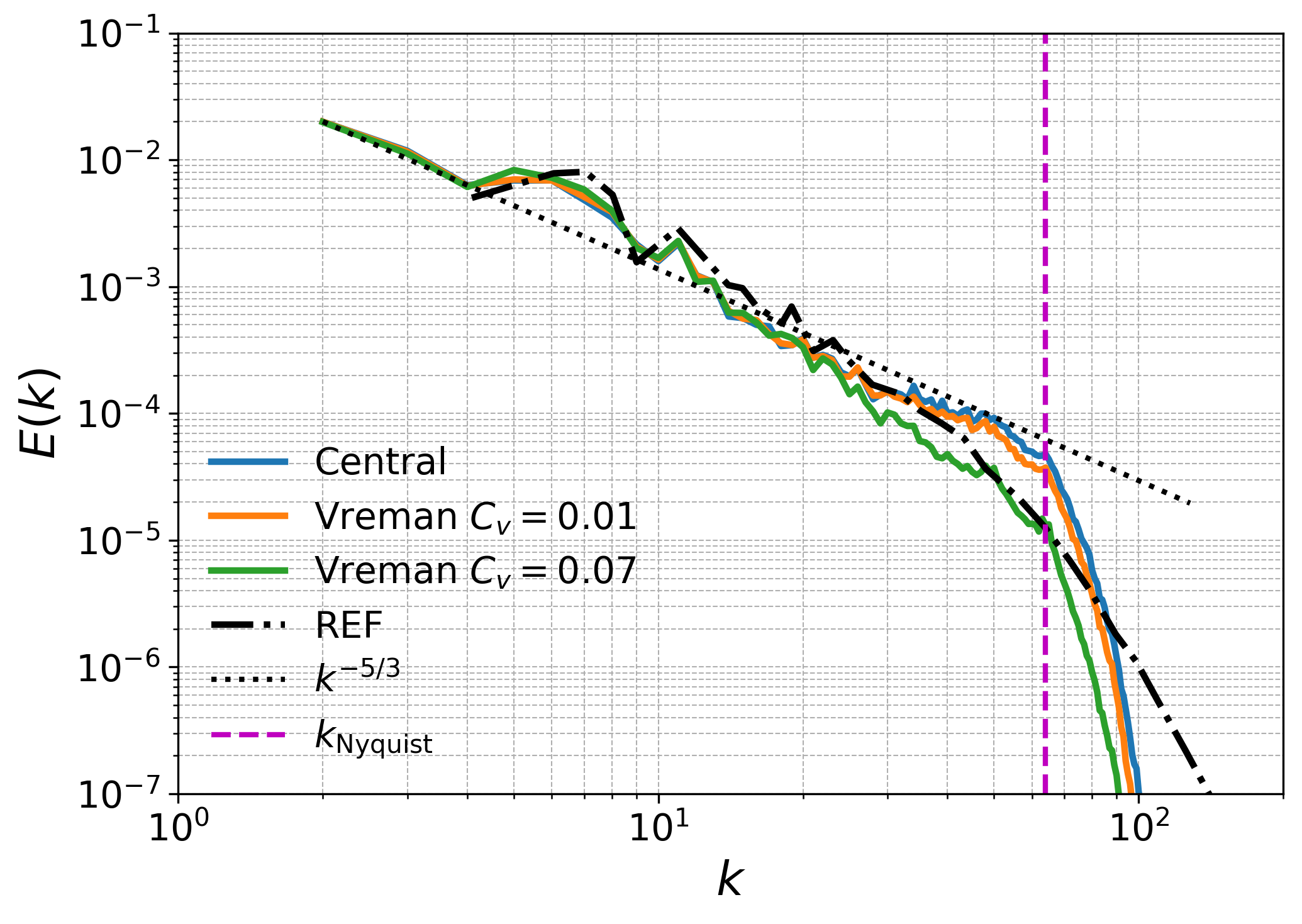}
        \caption{Kinetic energy spectrum at $t/t_c=9$.}
        \label{fig:part2_spectra}
    \end{subfigure}
    \caption{Effect of the Vreman SGS model on the Chandrasekar split form. The model does not improve the accuracy of the transitional dynamics, while in the fully turbulent regime larger constants recover the expected spectral decay at high wavenumbers. Increasing the constant shifts model-induced dissipation toward lower wavenumbers, overdamping intermediate scales, while smaller constants under-dissipate the highest wavenumbers.}
    \label{fig:part2}
\end{figure}

During the early transitional phase ($t/t_c \in [4,6]$), the larger constant ($C_v = 0.07$) introduces significant dissipation, reducing the dissipation peak, while $C_v = 0.01$ produces results nearly identical to the baseline case without SGS, only slightly mitigating the peak (Fig.~\ref{fig:part2_kinenrate}). In the spectral domain, $C_v = 0.07$ effectively suppresses energy accumulation at high wavenumbers ($k > 30$), but over-dissipates energy at intermediate scales and remains slightly insufficient near the cutoff range. Conversely, $C_v = 0.01$ does not provide adequate dissipation, and energy pile-up persists (Fig.~\ref{fig:part2_spectra}). 

\begin{figure}[h!]
    \centering
    \begin{subfigure}[t]{0.48\textwidth}
        \centering
        \includegraphics[width=\textwidth]{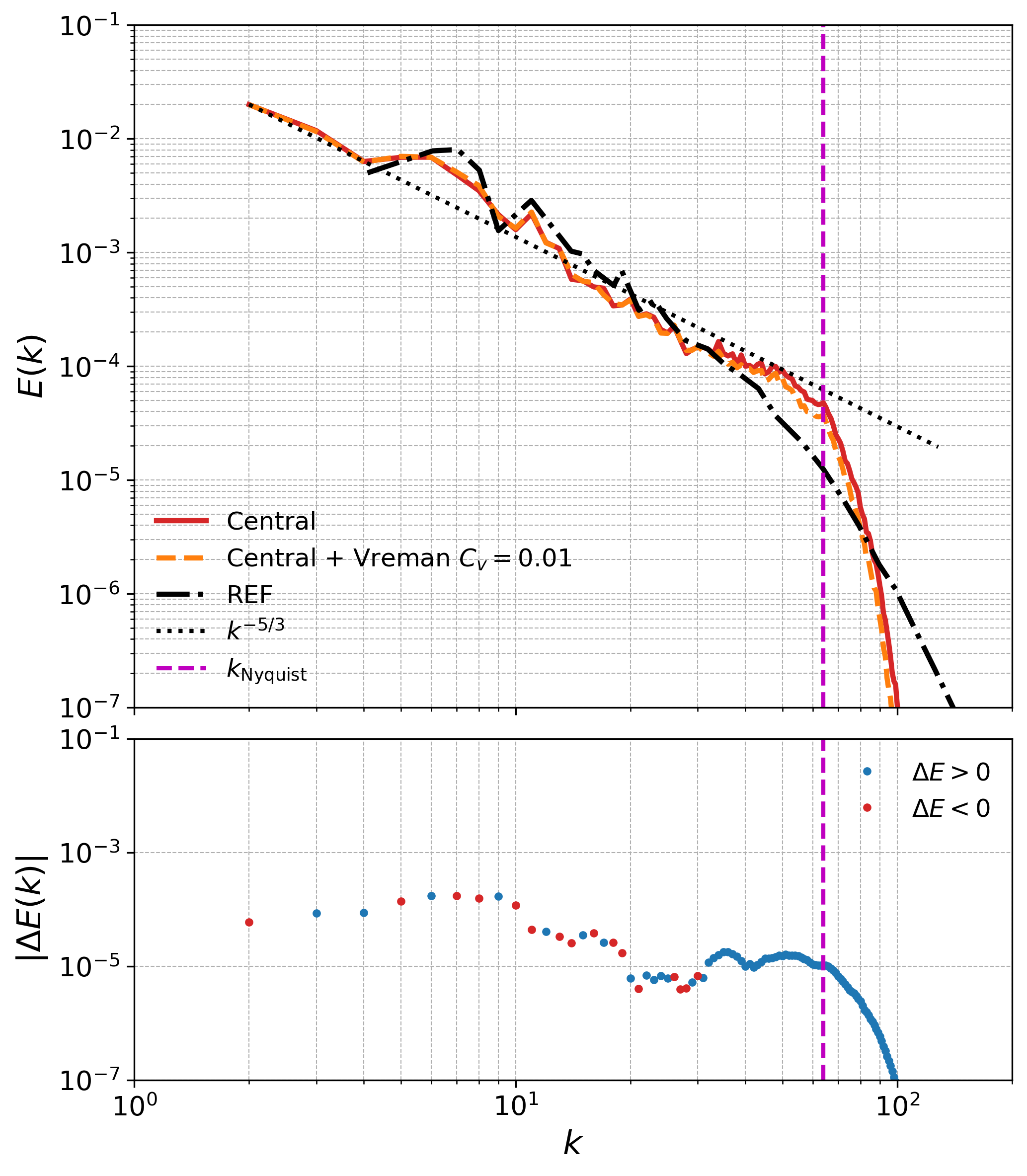}
        \caption{Vreman model with $C_v=0.01$.}
        \label{fig:Central_vs_CentralVreman001}
    \end{subfigure}
    \hfill
    \begin{subfigure}[t]{0.48\textwidth}
        \centering
        \includegraphics[width=\textwidth]{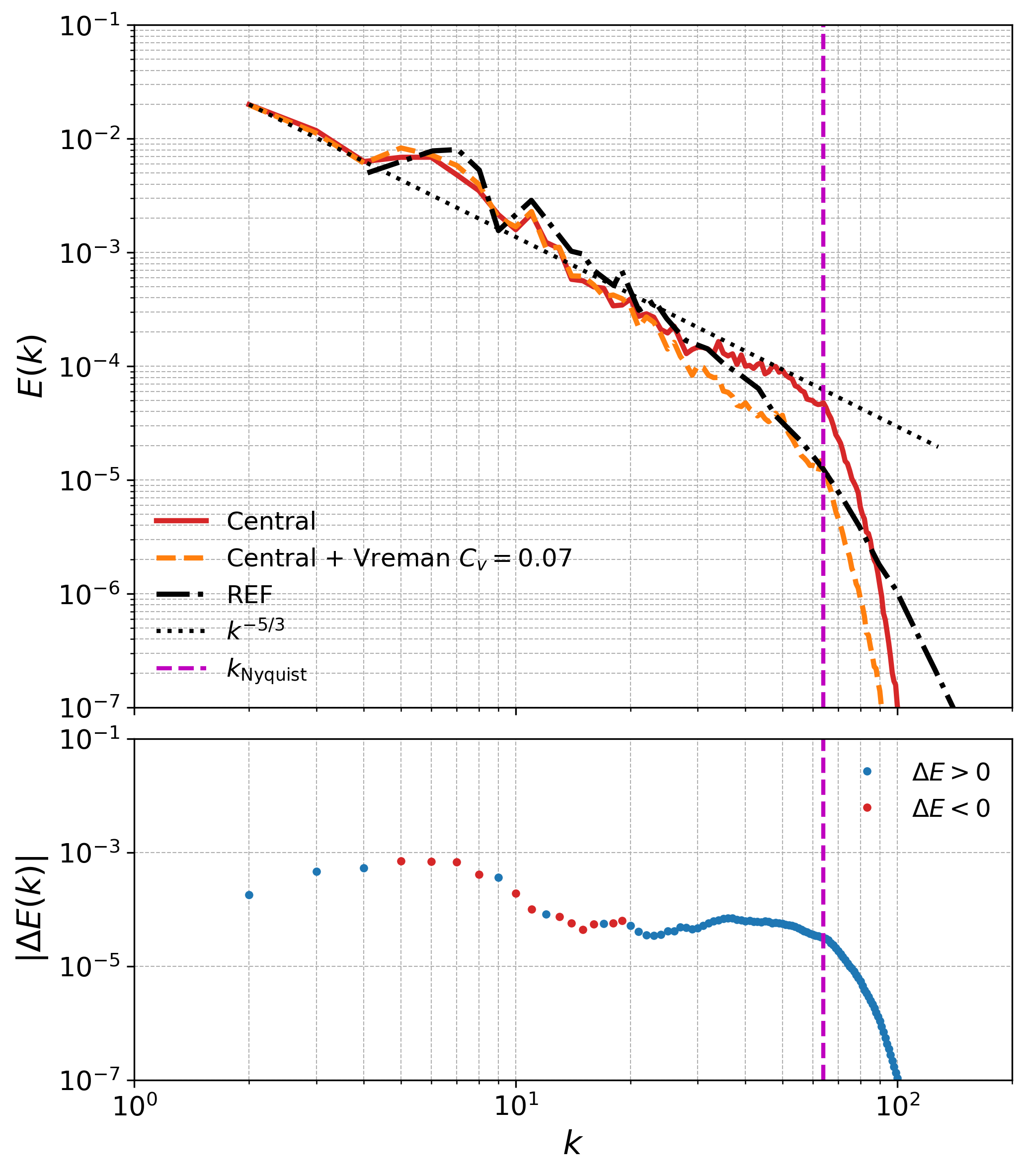}
        \caption{Vreman model with $C_v=0.07$.}
        \label{fig:Central_vs_CentralVreman007}
    \end{subfigure}
    \caption{Comparison between the baseline central scheme and the same scheme augmented with
    a Vreman SGS model of different strengths for the TGV at Reynolds 1600. Each subfigure shows the energy spectra (top)
    and the spectral difference $\Delta E(k)$ (bottom). The figures highlight how the SGS model constant controls both the magnitude of the added dissipation and the wavenumber at which it becomes active.}
    \label{fig:sgs_appendix_central_1600} 
\end{figure}

To examine this phenomenon in greater detail, Fig.~\ref{fig:sgs_appendix_central_1600} is presented. 
Each subplot in the figure contains two panels: the upper panel displays the energy spectra \( E(k) \) for both configurations on a logarithmic scale, while the lower panel shows the spectral difference
\[
\Delta E(k) = E_{\mathrm{base}}(k) - E_{\mathrm{SGS}}(k),
\]
which quantifies the modification of the energy content across scales induced by the SGS model. 
Since the results are represented on logarithmic scales, the absolute value of the spectral difference is plotted in the lower panel. 
To retain information on the sign of \(\Delta E(k)\), two reference lines are added. 
It should be noted that the simultaneous presence of positive and negative values indicates that the subgrid model has a negligible effect, with values oscillating around zero. 
Conversely, consistently positive values reveal that the SGS model drains energy from the resolved scales, allowing one to identify the wavenumber at which this energy removal becomes effective.

In both cases, the difference between simulations with and without the SGS model exhibits a similar trend. Up to a certain wavenumber, both spectra remain nearly identical, with the difference oscillating around zero. The magnitude of this difference scales with the model constant and remains much smaller than the flow energy at those wavenumbers, which explains its absence in the upper figures. For the weaker model, this transition occurs around \( k \approx 30 \), and for the stronger one, it occurs around \( k \approx 20 \). The model constant thus appears to control the range of resolved scales before the SGS model becomes active. Beyond this point, the turbulence model dissipates the energy accumulation near the cutoff wavenumber. The stronger model achieves this more effectively.

In summary, the inclusion of the Vreman SGS model does not enhance the transitional flow dynamics, and the solution quality in this regime remains limited. 
Once turbulence is developed, a larger model constant helps to achieve the expected spectral behavior. 
The intensity of dissipation at high wavenumbers depends on the specific wavenumber at which the model-induced dissipation becomes apparent. 
Consequently, intermediate scales tend to be over-damped when a large constant is used, while a smaller constant leaves the highest wavenumbers slightly under-dissipated. \textcolor{black}{Ideally, one would seek a dissipation profile that becomes active closer to \(k_{\mathrm{eff}}\), while still providing sufficient intensity to remove the excess energy near the cutoff.}

\subsubsection{Influence of Riemann Solvers}
Riemann solvers introduce upwind dissipation through characteristic decomposition, producing an interesting iLES behavior. In this part, the Chandrasekar split form is employed without explicit SGS modeling to isolate the effects of the Riemann solver choice.

\begin{figure}[h!]
    \centering
    \begin{subfigure}[t]{0.49\textwidth}
        \includegraphics[width=\textwidth]{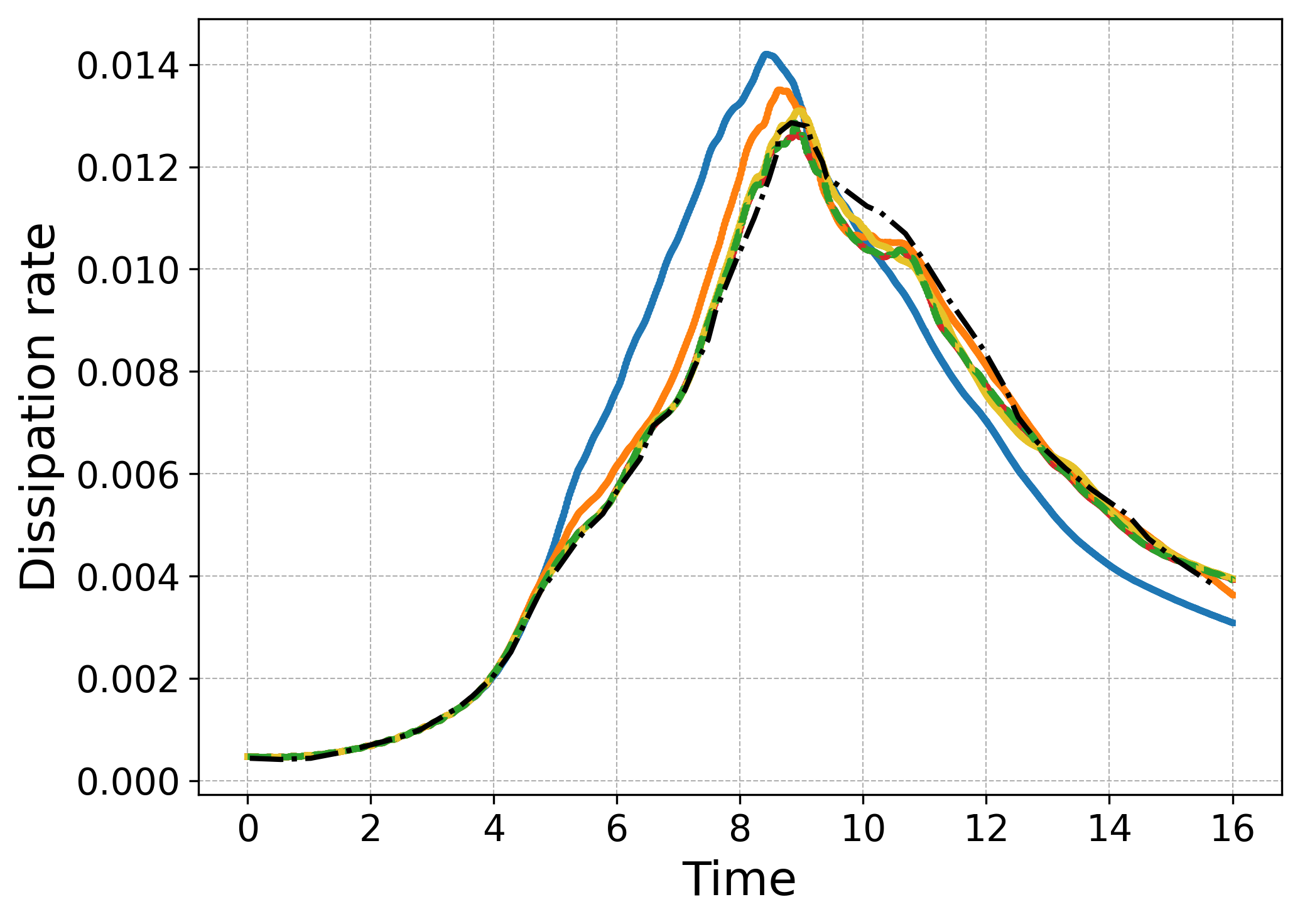}
        \caption{Kinetic energy dissipation rate.}
        \label{fig:part3_kinenrate}
    \end{subfigure}
    \hfill
    \begin{subfigure}[t]{0.49\textwidth}
        \includegraphics[width=\textwidth]{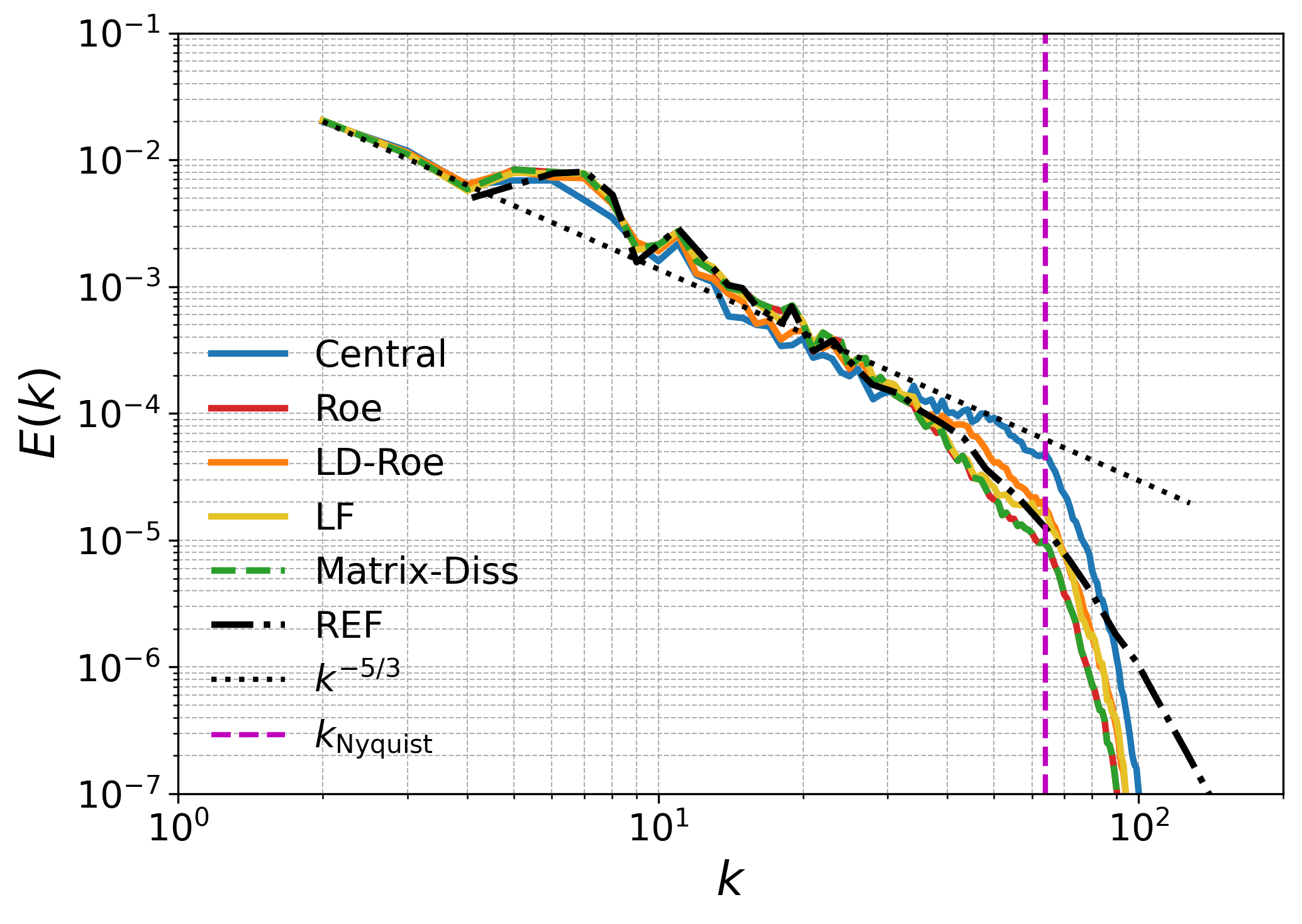}
        \caption{Kinetic energy spectrum at $t/t_c=9$.}
        \label{fig:part3_spectra}
    \end{subfigure}
    \caption{Effect of Riemann solver choice on dissipation and spectral behavior using the Chandrasekar split form without explicit SGS modeling. Roe, Matrix Dissipation, and Lax--Friedrichs supply controlled high-wavenumber dissipation that stabilizes the transition and improves agreement with the reference solution, while central and LD-Roe fluxes remain under-dissipative in the turbulent regime.} 
    \label{fig:part3}
\end{figure}

During the transitional phase ($t/t_c \in [4,8]$), Riemann solvers improve solution stability and accuracy. Central fluxes alone do not provide sufficient dissipation, and the split form compensates by introducing excessive implicit dissipation. The Roe and Matrix Dissipation fluxes yield smooth dissipation profiles that agree well with the reference solution. 
This observation is consistent with the findings of \cite{Manzanero2020SVV}, who reported that Riemann solvers primarily influence the high-wavenumber dynamics.

The low-dissipation Roe (LD-Roe) solver provides an intermediate approach: it partially relies on the split form for early stabilization but ultimately introduces more dissipation than optimal (Fig.~\ref{fig:part3_kinenrate}). In the spectral domain (Fig.~\ref{fig:part3_spectra}), all schemes except the less dissipative ones (central and LD-Roe) accurately capture the energy spectrum up to \( k \approx 30 \), while these flux formulations exhibit a slight loss of precision.
For larger wavenumbers, the central flux exhibits strong energy pile-up; LD-Roe exhibits a minor energy pile-up, while Roe and Matrix Dissipation slightly over-dissipate the highest wavenumbers, \textcolor{black}{which helps control energy accumulation close to the cutoff.} The Lax–Friedrichs scheme performs similarly to Roe but differs in that it shows a noticeable energy accumulation near the cutoff due to its over-upwind character at low Mach numbers, as previously reported in \cite{2017:Moura, WINTERS20181}.


\textcolor{black}{For the subsequent analysis, the Roe scheme and its low-dissipation variant, LD-Roe, are retained as representative upwind configurations to assess whether they can be fine-tuned by introducing a controlled amount of dissipation through the SGS model.}

\subsubsection{Hybrid iLES--SGS Configurations}
Hybrid configurations combining Riemann solvers with a weak Vreman SGS model ($C_v = 0.01$) are evaluated to determine whether a small explicit viscosity improves transitional and turbulent dynamics.

\begin{figure}[h!]
    \centering
    \begin{subfigure}[t]{0.49\textwidth}
        \includegraphics[width=\textwidth]{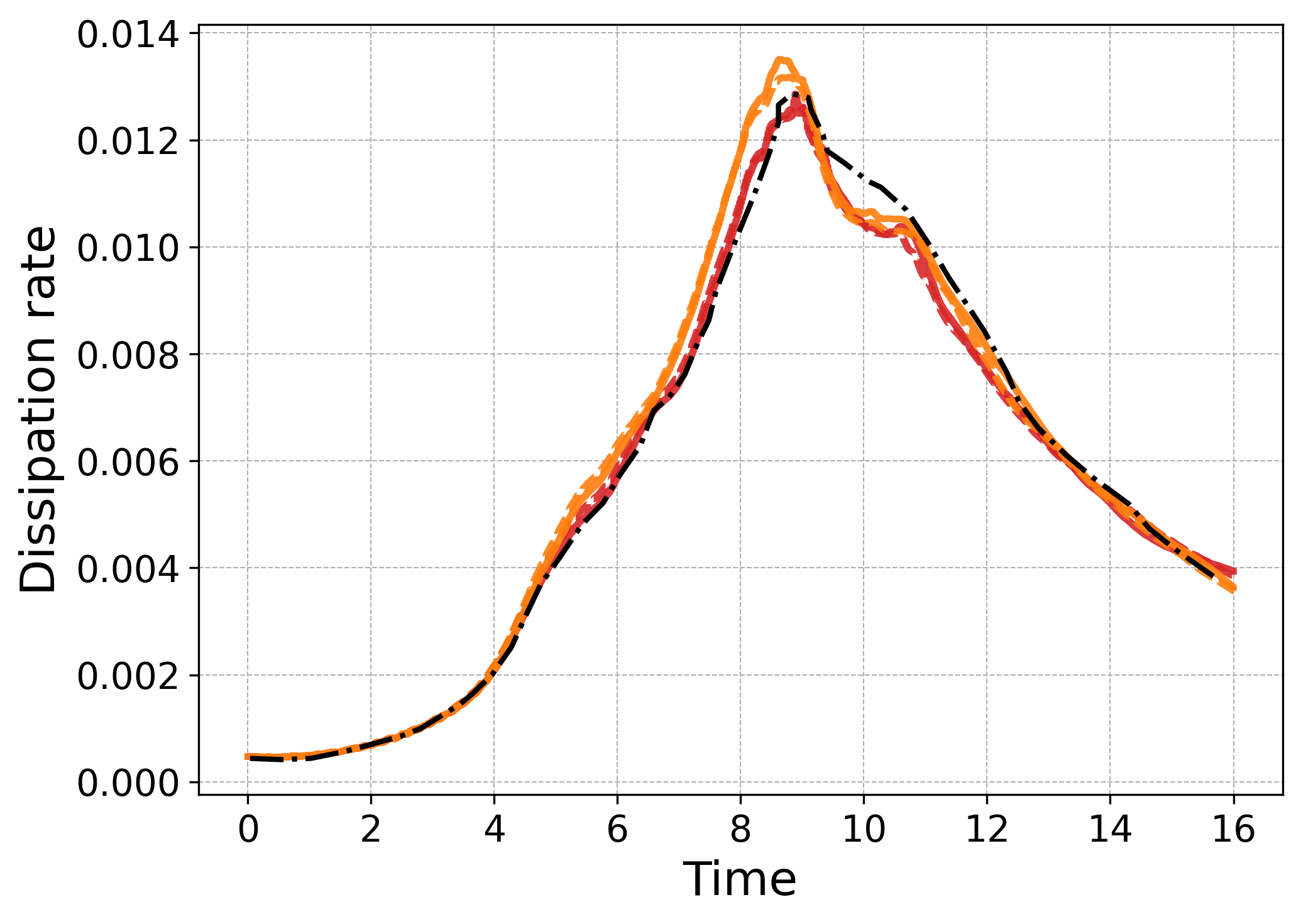}
        \caption{Kinetic energy dissipation rate.}
        \label{fig:part4_kinenrate}
    \end{subfigure}
    \hfill
    \begin{subfigure}[t]{0.49\textwidth}
        \includegraphics[width=\textwidth]{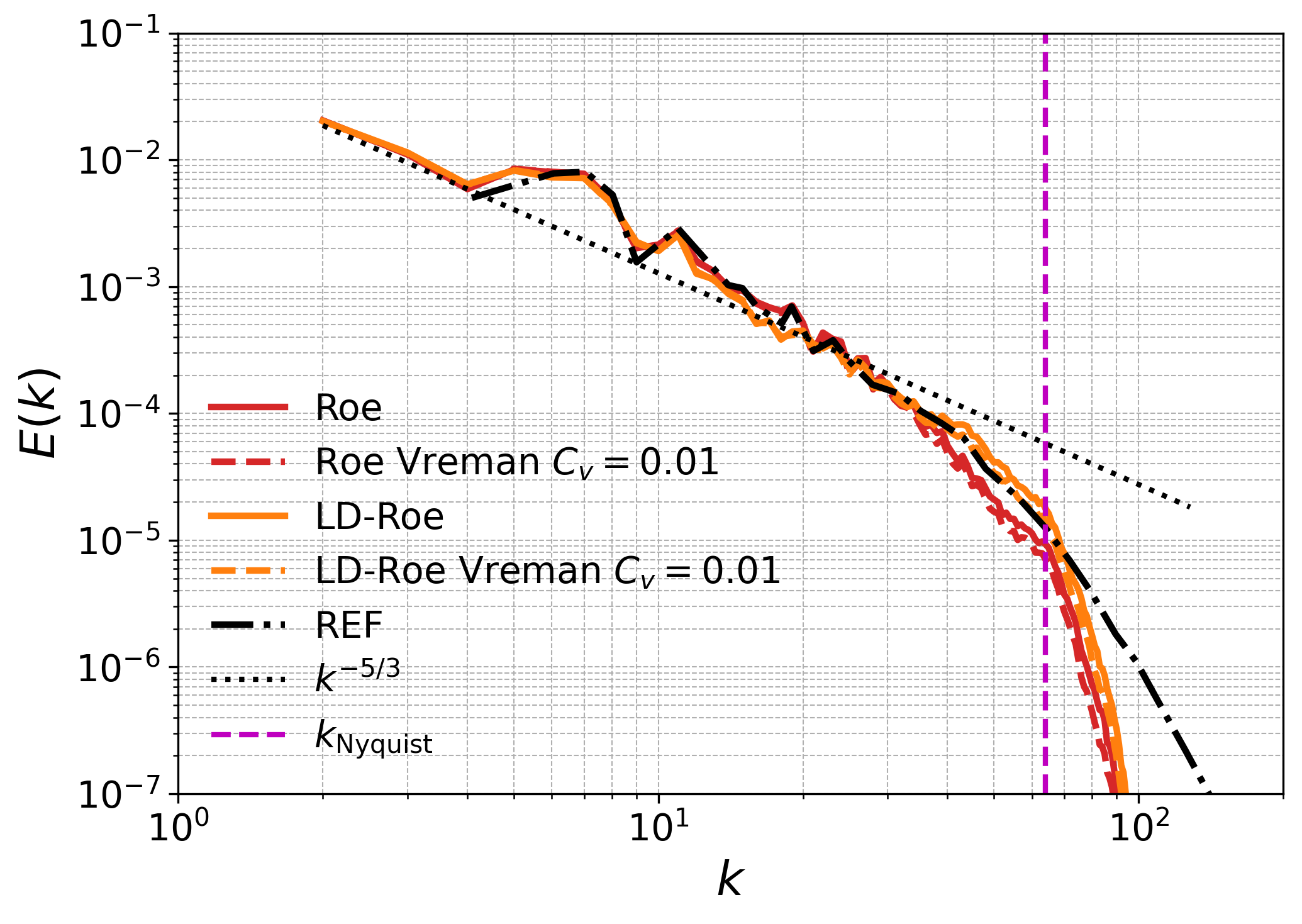}
        \caption{Kinetic energy spectrum at $t/t_c=9$.}
        \label{fig:part4_spectra}
    \end{subfigure}
    \caption{Effect of combining Riemann solvers with a weak Vreman SGS model ($C_v = 0.01$) on dissipation and spectral behavior. The weak SGS contribution slightly increases dissipation during transition, most noticeably for the LD-Roe--Vreman configuration, while the Roe flux remains largely unchanged. In the turbulent regime, LD-Roe coupled with the SGS model better mitigates energy pile-up at high wavenumbers, providing improved spectral fidelity compared to the purely iLES configurations.}

    \label{fig:part4}
\end{figure}

With the low Vreman constant ($C_v = 0.01$), the explicit subgrid-scale contribution is minor. During the transitional phase ($t/t_c < 8$), a slight increase in dissipation is observed, most notably for the LD-Roe--Vreman configuration, which exhibits a marginally reduced dissipation peak. In contrast, the standard Roe flux shows negligible change, indicating that the iLES dissipation from the Riemann solver dominates.

Beyond $t/t_c = 8$, the simulations without Vreman maintain slightly elevated dissipation, subtly damping small-scale energy. Spectrally (Fig.~\ref{fig:part4_spectra}), all configurations behave similarly for well-resolved scales ($k < 30$). For higher wavenumbers ($k > 30$), LD-Roe--Vreman shows improved spectral agreement with reference, effectively mitigating energy pile-up at the smallest resolved scales.

In summary, the weak Vreman SGS model has a limited impact during the transitional phase, offering only marginal benefits when combined with LD-Roe. \textcolor{black}{In this regime, the Roe flux provides the closest agreement with the reference among the configurations tested. Once turbulence is fully developed, Roe remains slightly over-dissipative at high wavenumbers, which helps control energy accumulation near the cutoff. By contrast, the LD-Roe scheme coupled with the Vreman model is slightly less dissipative and follows the reference spectrum more closely up to the near-cutoff range.}


\subsection{Under-Resolved LES (The Inviscid Taylor–Green Vortex): Necessity and Limitations of Explicit SGS Models}
\label{sec:InviscidTGV}






\textcolor{black}{In the previous section, it was shown that the weak Vreman SGS model has a minor influence during transition, with Roe-type configurations remaining closest to the reference among the cases tested.
In the turbulent regime, Roe becomes slightly over-dissipative, while LD-Roe combined with Vreman improves the high-wavenumber spectral agreement among the tested hybrid configurations.}
To further assess the robustness and dissipation properties of these configurations, we now consider the inviscid Taylor–Green vortex (Euler limit). 
This case removes physical viscosity, thereby isolating the numerical dissipation mechanisms—implicit, through the Riemann solver and the Chandrasekar split form, and explicit, through the Vreman model—in stabilizing the flow and sustaining the energy cascade. 
The objective is to determine whether the same trends persist under inviscid conditions.

\subsubsection{Hybrid iLES--SGS Configurations}
\begin{figure}[h!]
    \centering
    \begin{subfigure}[t]{0.49\textwidth}
        \includegraphics[width=\textwidth]{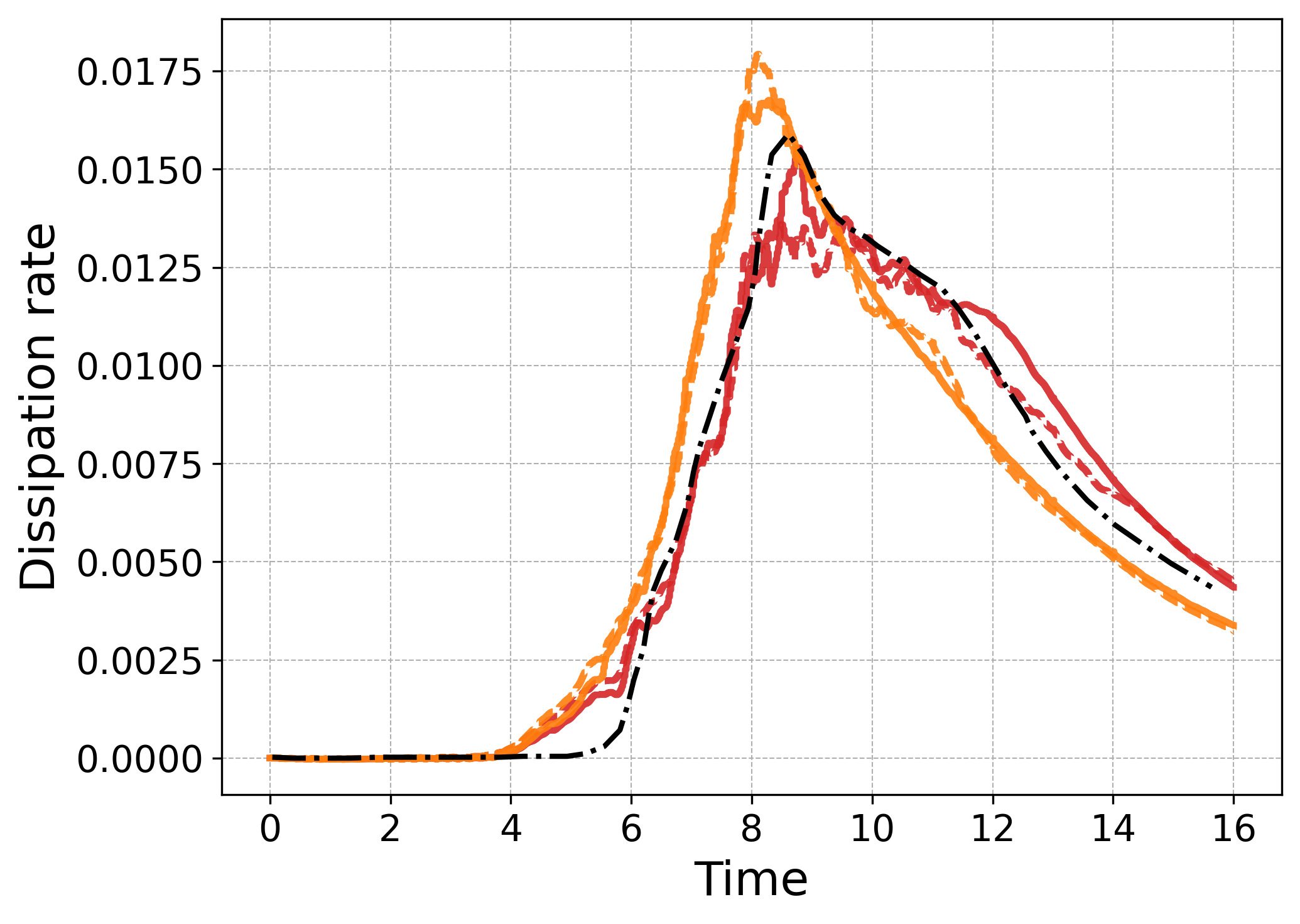}
        \caption{Kinetic energy dissipation rate.}
        \label{fig:part5_kinenrate}
    \end{subfigure}
    \hfill
    \begin{subfigure}[t]{0.49\textwidth}
        \includegraphics[width=\textwidth]{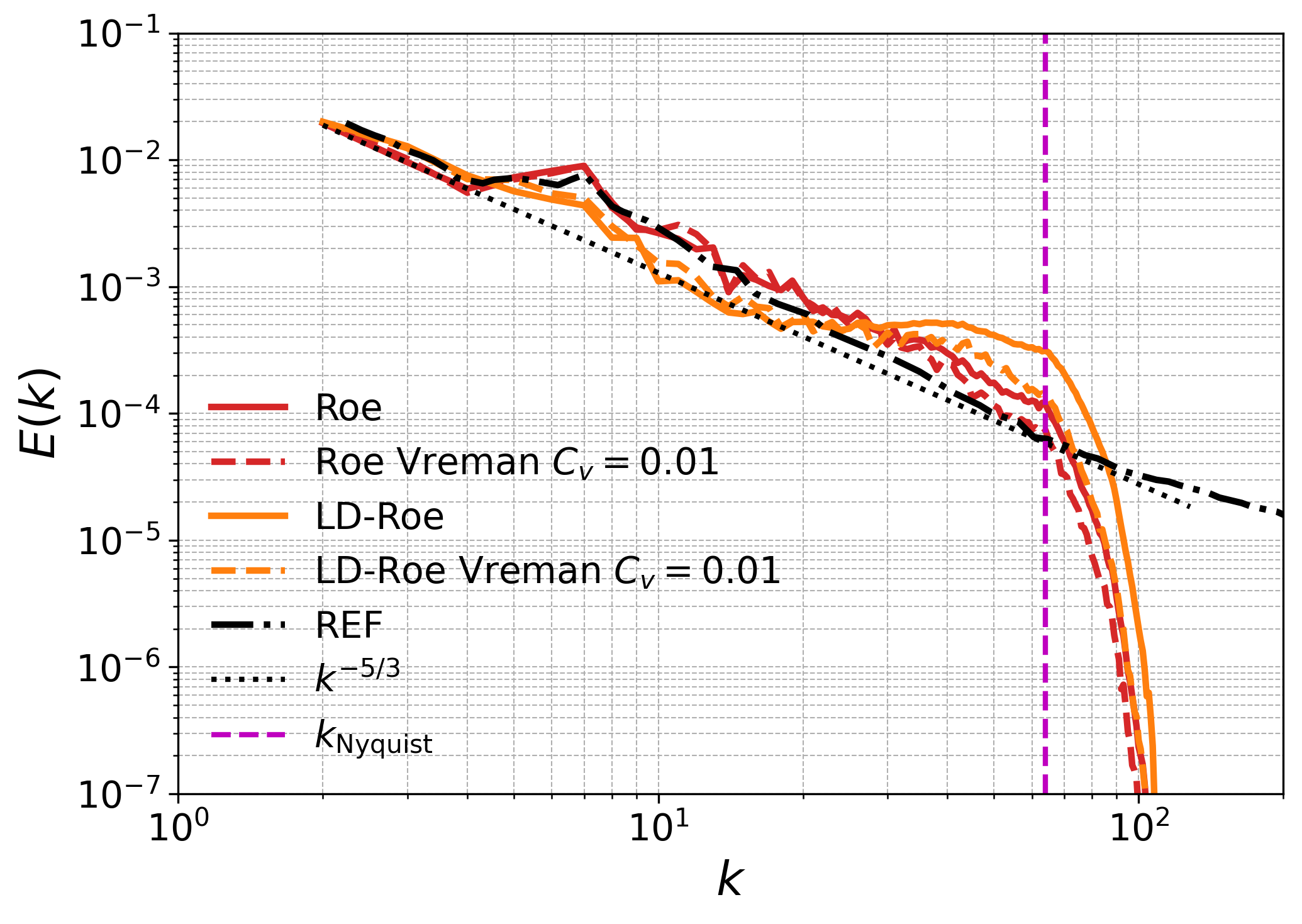}
        \caption{Kinetic energy spectrum at $t/t_c=9$.}
        \label{fig:part5_spectra}
    \end{subfigure}
    \caption{Inviscid TGV: hybrid iLES--SGS configurations combining Riemann solvers with a weak Vreman model ($C_v = 0.01$). \textcolor{black}{During transition, Roe with and without SGS show very similar trends and remain close to the high-resolution reference, whereas LD-Roe exhibits increased dissipation associated with split-form activation. In the turbulent regime, the weak Vreman model improves the high-wavenumber behavior for both fluxes, with its main effect concentrated at small scales.}}
    \label{fig:part5}
\end{figure}

We repeat the analysis from the previous section, combining the Riemann solvers with a weak Vreman SGS model (\( C_v = 0.01 \)) and show the results in Fig.~\ref{fig:part5}. 

During the laminar phase (\( t/t_c \lesssim 5 \)), all configurations exhibit similar behavior with minimal dissipation, and the weak Vreman SGS model has a negligible effect.

In the transitional regime ($t/t_c \gtrsim 5$), differences emerge. LD-Roe dissipates more energy due to the combination of insufficient built-in dissipation and the activation of the split form, while Roe remains closer to the high-resolution reference. 

\textcolor{black}{ The effect of the Vreman model is mild and depends on the numerical flux. With Roe, adding the Vreman model slightly reduces the dissipation peak, whereas with LD-Roe it slightly increases it. Overall, Roe simulations with and without the Vreman model exhibit very similar behavior during transition, indicating that the SGS contribution considered here does not significantly modify the transitional energy-dissipation dynamics when the Roe flux is used.}

In the fully developed turbulent phase, spectral behavior shows that Roe maintains energy levels reasonably well up to $k \approx 20$, while LD-Roe underestimates energy in this range and exhibits a high-wavenumber pile-up for $k > 20$. Adding Vreman to LD-Roe mitigates, but does not completely remove, the pile-up. Roe--Vreman introduces extra dissipation for $k > 30$, improving high-wavenumber behavior. 

\textcolor{black}{In summary, Roe gives the closest overall agreement among the tested configurations across the transitional and turbulent stages of the inviscid TGV.}
The inclusion of a weak Vreman SGS model enhances dissipation at high wavenumbers, improving the turbulent regime but slightly reducing accuracy during transition. 
Notably, this conclusion differs from that of the viscous TGV case. 
In the following section, we refine the SGS constant to assess whether a higher value yields improved performance under inviscid conditions.

\subsubsection{Fine-Tuning of the SGS Constant}
\label{sec:SGSenergy}
In this section, we examine in detail the influence of the Vreman constant on the accuracy of the simulations. We retain the Roe flux, which exhibited the most consistent performance across the different regimes, and assess the impact of incorporating the Vreman model using two values of the constant: \(C_v = 0.01\) and \(C_v = 0.07\).

\begin{figure}[h!]
    \centering
    \begin{subfigure}[t]{0.49\textwidth}
        \includegraphics[width=\textwidth]{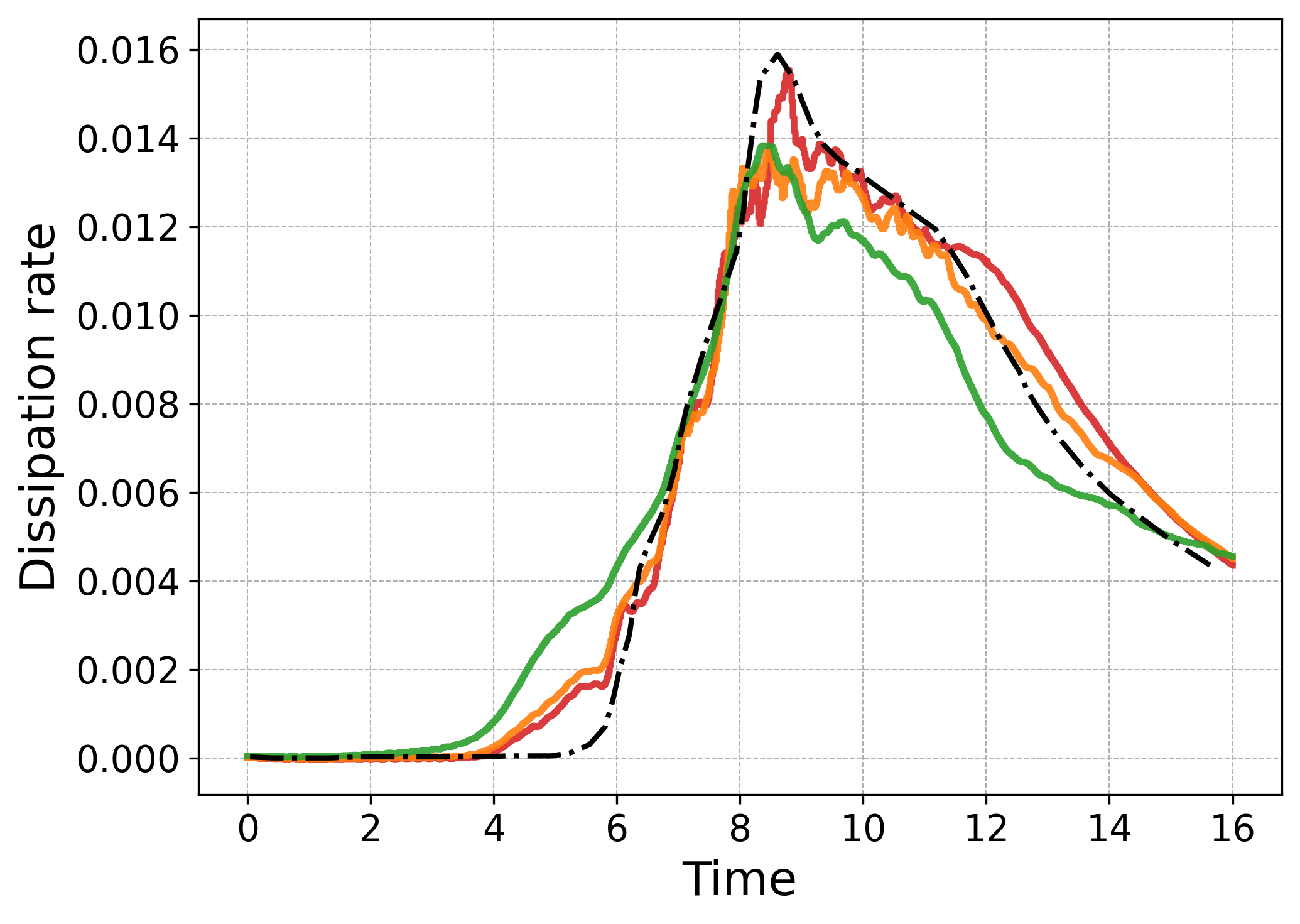}
        \caption{Kinetic energy dissipation rate.}
        \label{fig:part6_kinenrate}
    \end{subfigure}
    \hfill
    \begin{subfigure}[t]{0.49\textwidth}
        \includegraphics[width=\textwidth]{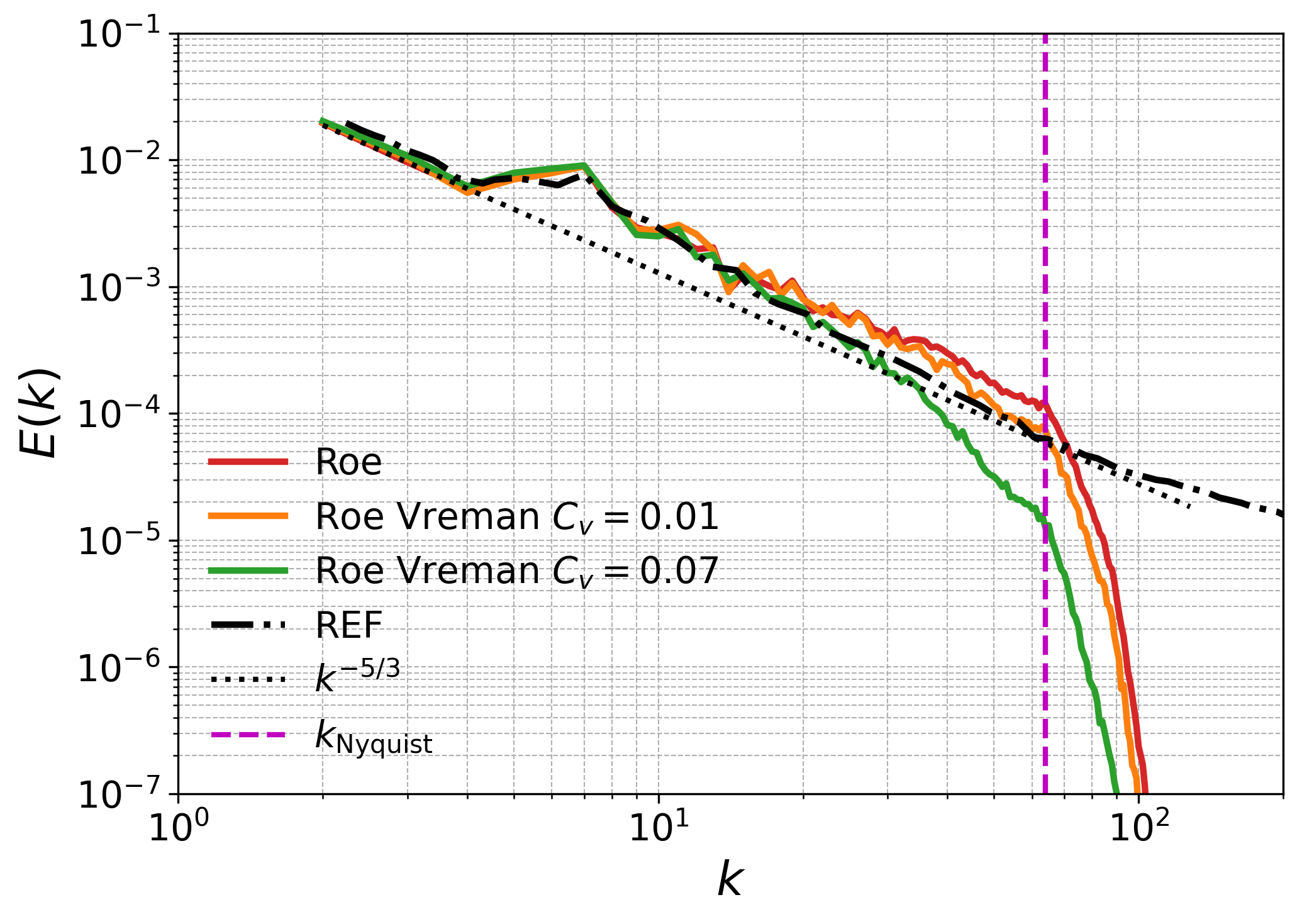}
        \caption{Kinetic energy spectrum at $t/t_c=9$.}
        \label{fig:part6_spectra}
    \end{subfigure}
    \caption{Inviscid TGV: Effect of the Vreman SGS constant on dissipation and spectral behavior for hybrid iLES--SGS configurations with the Roe flux. During transition, all SGS configurations reduce accuracy compared to the baseline Roe scheme. In the turbulent regime, a small constant ($C_v = 0.01$) mitigates high-wavenumber energy pile-up but overestimates energy at intermediate wavenumbers, while a larger constant ($C_v = 0.07$) improves dissipation around $k \approx 20$ but over-damps the highest wavenumbers.}
    \label{fig:part6}
\end{figure}

As observed previously, incorporating the Vreman model does not enhance performance during the transition phase. 
As shown in Fig.~\ref{fig:part6_kinenrate}, a small amount of Vreman dissipation (\( C_v = 0.01 \)) already reduces accuracy, while increasing it to \( C_v = 0.07 \) further degrades the solution.

Focusing now on the energy spectra (Fig.~\ref{fig:part6_spectra}), the Roe scheme retains too much energy compared to the reference solution. Increasing the constant to $C_v = 0.01$ allows matching the reference near the cutoff, but leads to excessive energy between $k = 20$ and the cutoff wavenumber. In contrast, increasing the constant to the typical finite-volume value of $C_v = 0.07$ aligns the spectrum around $k = 20$, but results in over-dissipation for $k > 20$. 

\begin{figure}[h!]
    \centering
    \begin{subfigure}[t]{0.48\textwidth}
        \centering
        \includegraphics[width=\textwidth]{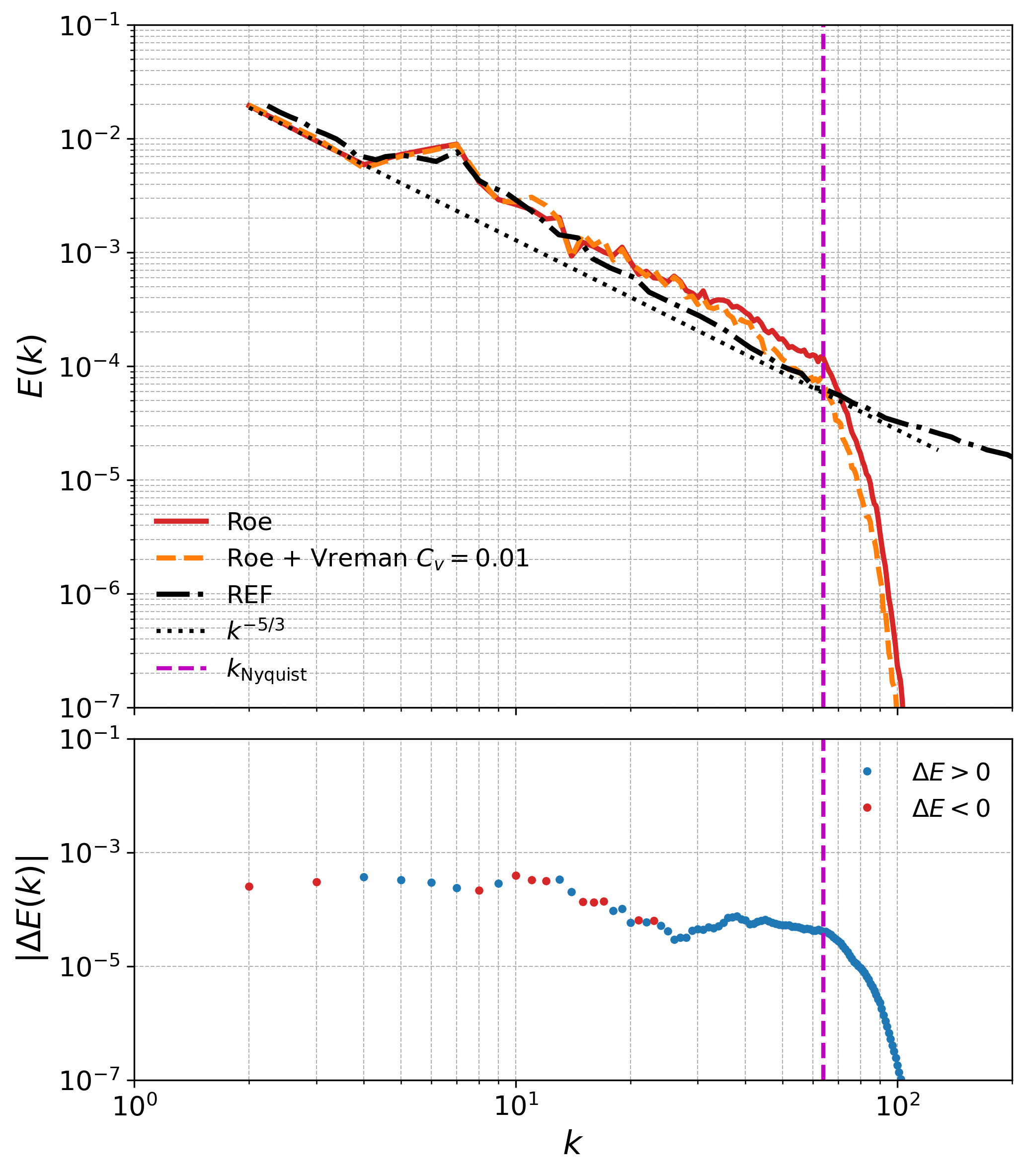}
        \caption{Vreman model with $C_v=0.01$.}
        \label{fig:Roe_vs_RoeVreman001}
    \end{subfigure}
    \hfill
    \begin{subfigure}[t]{0.48\textwidth}
        \centering
        \includegraphics[width=\textwidth]{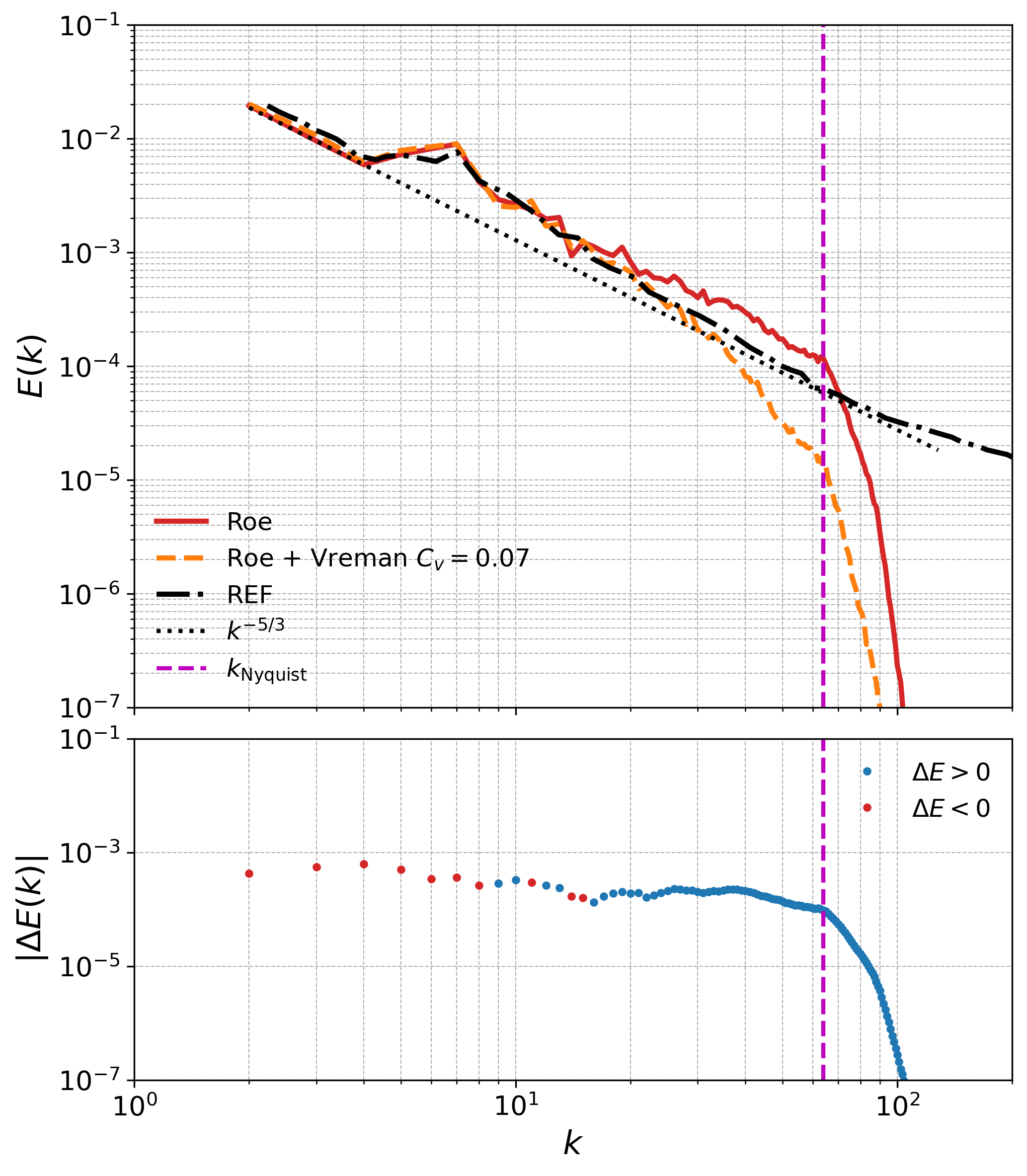}
        \caption{Vreman model with $C_v=0.07$.}
        \label{fig:Roe_vs_RoeVreman007}
    \end{subfigure}
    \caption{Comparison between the baseline Roe scheme and the same scheme augmented with
    a Vreman SGS model of different strengths for the inviscid TGV. Each subfigure shows the energy spectra (top)
    and the spectral difference $\Delta E(k)$ (bottom). The figures highlight how the SGS model constant controls both the magnitude of the added dissipation and the wavenumber at which it becomes active.}
    \label{fig:sgs_appendix_roe_inv}
\end{figure}

To examine this phenomenon in greater detail, Fig.~\ref{fig:sgs_appendix_roe_inv}  quantifies the modification of the energy content across scales induced by the SGS model. 
Once again, the model with the lower constant adds dissipation at higher wavenumbers (compared to the model with the larger constant). The dissipation profiles appear flatter, likely due to the under-resolved (inviscid) nature of this test case. The profile is noticeably flatter for the higher-constant case. 
\textcolor{black}{The higher constant provides the level of dissipation needed to control the excess of energy, but it acts too broadly across the spectrum. Ideally, one would retain the stronger dissipation associated with \(C_v=0.07\), while concentrating it only in the high-wavenumber range, closer to the cutoff, so that intermediate scales are not excessively damped.}

\textcolor{black}{\subsection{Quantitative spectral-error assessment}}
\label{sec:spectral_error_assessment}
To complement the visual comparison of energy spectra, we compute band-integrated spectral errors for selected representative configurations. The objective is not to define a new optimization criterion, but to quantify the spectral trade-offs discussed above, namely overdamping of resolved and intermediate scales and energy accumulation near the cutoff range.

For each wavenumber band \(B\), the normalized absolute spectral error is defined as
\begin{equation}
\mathcal{E}_{B}
=
\frac{
\int_{B} \left| E(k)-E_{\mathrm{ref}}(k) \right| \, \mathrm{d}k
}{
\int_{B} E_{\mathrm{ref}}(k) \, \mathrm{d}k
},
\end{equation}
and the signed spectral bias as
\begin{equation}
\mathcal{S}_{B}
=
\frac{
\int_{B} \left( E(k)-E_{\mathrm{ref}}(k) \right) \, \mathrm{d}k
}{
\int_{B} E_{\mathrm{ref}}(k) \, \mathrm{d}k
}.
\end{equation}
Here, \(E(k)\) is the kinetic-energy spectrum of the simulation and \(E_{\mathrm{ref}}(k)\) is the corresponding reference spectrum interpolated onto the same wavenumber grid. The quantity \(\mathcal{E}_{B}\) measures the magnitude of the spectral discrepancy in each band, whereas \(\mathcal{S}_{B}\) indicates its sign: positive values correspond to excess energy, while negative values indicate overdamping.

Since this diagnostic is intended to quantify the effect of dissipation mechanisms on the turbulent cascade and the cutoff range, band-integrated errors are evaluated only for wavenumbers \(k>10\). The largest energy-containing scales are therefore excluded from this metric, as they are comparatively less affected by the SGS model and by the high-wavenumber dissipation mechanisms considered here. Three spectral ranges are distinguished: an intermediate range, \(10 < k \le 30\), a high-wavenumber range, \(30 < k \le k_{\mathrm{eff}}\), and a near-Nyquist diagnostic range, \(k_{\mathrm{eff}} < k \le k_{\mathrm{Ny}}\). For the configurations considered here, the discretization has 128 degrees of freedom per direction, giving \(k_{\mathrm{eff}}\simeq 42.7\) and \(k_{\mathrm{Ny}}=64\). The intermediate and high-wavenumber bands are used to quantify the spectral agreement over the usable range of the polynomial approximation, whereas the near-Nyquist range is included only as a diagnostic of energy accumulation close to the nominal cutoff.

The selected configurations are organized into three groups. The first group assesses the effect of polynomial order in the well-resolved TGV at \(Re=1600\). This comparison confirms that the baseline iLES approach already provides accurate results in this regime. Adding the explicit SGS model does not improve the solution; instead, it increases the spectral error and makes the \(16P7\) result less accurate than the lower-order \(32P3\) configuration.

The other two groups focus on representative \(P=7\) configurations in the well-resolved TGV at \(Re=1600\) and in the strongly under-resolved inviscid TGV. These cases are selected to examine the interaction between the numerical dissipation supplied by the Riemann solver and the additional dissipation introduced by the Vreman SGS model.

Table~\ref{tab:band_spectral_errors} reports the resulting band-integrated errors. In the well-resolved polynomial-order comparison, the \(16P7\) iLES configuration has smaller errors than the corresponding \(32P3\) case, especially in the high-wavenumber band. When the standard Vreman coefficient, \(C_v=0.07\), is added to the \(16P7\) case, the error increases in all bands and the bias becomes strongly negative. This confirms that, in this regime, the explicit SGS contribution introduces excessive dissipation into a solution that is already sufficiently resolved.

\begin{table}[t]
\centering
\caption{Band-integrated spectral errors for representative configurations. The absolute error \(\mathcal{E}_B\) measures the magnitude of the spectral error, while the signed bias \(\mathcal{S}_B\) distinguishes energy excess (\(\mathcal{S}_B>0\)) from overdamping (\(\mathcal{S}_B<0\)). The near-Nyquist band is used only as a diagnostic of cutoff-scale behavior.}
\label{tab:band_spectral_errors}

{
\small
\setlength{\tabcolsep}{0pt}
\renewcommand{\arraystretch}{1.08}
\begin{tabular*}{\textwidth}{@{\extracolsep{\fill}}lrrrrrr@{}}
\toprule
\multirow{2}{*}{Configuration}
& \multicolumn{2}{c}{Mid}
& \multicolumn{2}{c}{High}
& \multicolumn{2}{c}{Near Nyquist} \\
\cmidrule(lr){2-3}\cmidrule(lr){4-5}\cmidrule(l){6-7}
& \(\mathcal{E}_B\) & \(\mathcal{S}_B\)
& \(\mathcal{E}_B\) & \(\mathcal{S}_B\)
& \(\mathcal{E}_B\) & \(\mathcal{S}_B\) \\
\midrule

\multicolumn{7}{@{}l}{\textit{Well-resolved polynomial-order comparison} (Fig.~\ref{fig:32P3_vs_16P7_spectra})} \\
\addlinespace[2pt]
\quad \(32P3\) 
& 0.115 &  0.031
& 0.449 & -0.449
& 0.558 & -0.558 \\
\quad \(16P7\) 
& 0.095 & -0.002
& 0.128 & -0.120
& 0.346 & -0.346 \\
\quad \(16P7,\ C_v=0.07\) 
& 0.170 & -0.165
& 0.588 & -0.588
& 0.809 & -0.809 \\

\addlinespace[5pt]
\multicolumn{7}{@{}l}{\textit{Well-resolved flux--SGS interaction} (Fig.~\ref{fig:part4_spectra})} \\
\addlinespace[2pt]
\quad LD-Roe 
& 0.179 & -0.160
& 0.062 &  0.034
& 0.344 &  0.344 \\
\quad LD-Roe, \(C_v=0.01\) 
& 0.169 & -0.162
& 0.086 & -0.080
& 0.080 &  0.080 \\
\quad Roe 
& 0.095 & -0.002
& 0.128 & -0.120
& 0.346 & -0.346 \\
\quad Roe, \(C_v=0.01\) 
& 0.082 & -0.028
& 0.226 & -0.222
& 0.454 & -0.454 \\

\addlinespace[5pt]
\multicolumn{7}{@{}l}{\textit{Strongly under-resolved SGS sensitivity} (Fig.~\ref{fig:part6_spectra})} \\
\addlinespace[2pt]
\quad Roe 
& 0.264 &  0.195
& 0.730 &  0.730
& 0.709 &  0.709 \\
\quad Roe, \(C_v=0.01\) 
& 0.356 &  0.293
& 0.441 &  0.441
& 0.158 &  0.155 \\
\quad Roe, \(C_v=0.07\) 
& 0.137 &  0.040
& 0.325 & -0.325
& 0.680 & -0.680 \\

\bottomrule
\end{tabular*}
}
\end{table}

The flux--SGS comparison further quantifies the dependence of the SGS effect on the baseline numerical dissipation. For LD-Roe, the weak Vreman model reduces the positive high- and near-Nyquist bias, indicating that it helps control energy accumulation close to the cutoff. The effect of the Vreman model in the mid-wavenumber range is negligible. 
For Roe, which already provides stronger high-wavenumber dissipation, the same weak SGS contribution shifts the high-wavenumber and near-Nyquist bands further toward negative bias. Therefore, the Vreman model can be beneficial or detrimental, depending on the dissipation of the baseline scheme.

Finally, in the strongly under-resolved inviscid configuration, Roe without SGS exhibits positive high-wavenumber and near-Nyquist biases, indicating energy accumulation. The weak Vreman model substantially reduces the near-Nyquist bias, but the bias remains positive in the high-wavenumber band. Increasing the coefficient to \(C_v=0.07\) further reduces the intermediate- and high-wavenumber absolute errors, but changes the sign of the high-wavenumber and near-Nyquist biases, indicating overdamping at those wavenumbers. 
A practical conclusion from these results is that, for the \(P=7\) configurations considered here, a weak Vreman contribution (\(C_v=0.01\)) provides a relatively non-intrusive form of additional dissipation. Its effect is minor in the laminar and transitional regimes, while in under-resolved turbulent regimes it can help reduce energy accumulation near the cutoff. However, it is not uniformly optimal across all spectral bands or Riemann solvers; in severely under-resolved cases, larger model constants may further reduce high-wavenumber errors, at the cost of possible overdamping.\\

\section{\textcolor{black}{Conclusions} }
\label{sec:conclusions}

This work examined the dissipation balance in high-order DGSEM simulations of the Taylor--Green vortex at \(Re=1600\) and in the inviscid limit. The objective was to assess how explicit Vreman SGS modeling interacts with the dissipation already introduced by split-form stabilization and Riemann solvers. The test matrix considered different polynomial orders, spatial resolutions, Riemann fluxes, and Vreman model constants, with particular attention to very high-order configurations motivated by modern GPU-oriented implementations.

The results confirm that the effect of explicit SGS modeling cannot be assessed independently of the underlying numerical dissipation. At comparable degrees of freedom, increasing the polynomial order improves the representation of the resolved scales, but also reduces the inherent damping near the cutoff range. This makes very high-order simulations more sensitive to the combined action of the split form, the Riemann flux, and the SGS model. The GPU performance results provide a practical motivation for considering such high polynomial orders, but the dissipation-balance issue itself arises from the numerical and modeling interaction, not from the hardware architecture.

For the well-resolved \(Re=1600\) configurations considered here, the iLES approach provides close agreement with the reference solution. In these cases, the dissipation supplied by the split form and the Riemann solver is sufficient to control the smallest resolved scales, and the addition of a Vreman model does not improve the solution. In particular, the standard coefficient \(C_v=0.07\) introduces excessive dissipation and degrades the energy spectra, while the weaker value \(C_v=0.01\) has only a limited effect.

The under-resolved cases show a different balance. When numerical dissipation alone is insufficient, high-wavenumber energy accumulation appears near the cutoff range, especially for very high-order discretizations. In this regime, an explicit SGS contribution can improve the spectral behavior, but its strength must be chosen carefully. A weak Vreman model provides a less intrusive correction, whereas a larger coefficient can remove more excess energy in severely under-resolved configurations but may also overdamp part of the resolved spectrum. Thus, the appropriate SGS strength depends on the resolution regime and on the dissipation already introduced by the numerical flux.

One practically relevant finding of this work is that, within the present test matrix, the Chandrasekar split form combined with the Roe flux provides a robust baseline for very high polynomial order, \(P=7\), in the laminar and transitional regimes. Adding a weak Vreman model, \(C_v=0.01\), has only a minor effect in these regimes and does not substantially alter the corresponding energy-dissipation dynamics. In under-resolved turbulent regimes, the same weak SGS contribution can help reduce energy accumulation near the cutoff, although its effect depends on the baseline Riemann-solver dissipation and is not uniformly optimal across all spectral bands. Stronger model constants may therefore still be advisable in severely under-resolved LES configurations, at the cost of possible overdamping.

The detailed comparisons also show that the SGS model acts in conjunction with the Riemann solver rather than as an independent correction. For less dissipative fluxes, such as LD-Roe, a weak Vreman contribution can help reduce high-wavenumber energy accumulation. For more dissipative Roe-type fluxes, the same SGS contribution may shift the high-wavenumber range toward overdamping. The band-integrated spectral-error analysis supports this interpretation by distinguishing configurations dominated by energy excess from those dominated by excessive dissipation.

Overall, no single combination of split form, Riemann solver, and Vreman constant provides the closest agreement across all regimes considered. Explicit SGS modeling is neutral or detrimental in well-resolved laminar and transitional regimes, but can become beneficial when under-resolution produces high-wavenumber energy accumulation. These results motivate scale-aware stabilization strategies in which numerical and modeled dissipation are adjusted according to polynomial order, resolution level, flow regime, and local spectral content.

The present study is limited to the Taylor--Green vortex, which provides a controlled setting for isolating dissipation mechanisms. Future work will extend the analysis to wall-bounded flows, where near-wall resolution, SGS modeling, and numerical dissipation interact differently. Further developments will also consider dynamically tuned SGS constants and modified SGS or spectral-viscosity formulations capable of targeting the cutoff range without unnecessarily damping the resolved turbulent cascade.\\

\section{Acknowledgements}
GR, GN, MCM, EV and EF acknowledge the funding received by the Grant DeepCFD (Project No. PID2022-137899OB-I00) funded by MICIU/AEI/10.13039/501100011033 and by ERDF, EU.
This research has received funding from the European Union (ROSAS, project number 101138319).
This research has received funding from the European Union (ERC, Off-coustics, project number 101086075). Views and opinions expressed are, however, those of the authors only and do not necessarily reflect those of the European Union or the European Research Council. Neither the European Union nor the granting authority can be held responsible for them.
We thankfully acknowledge the computer resources at MareNostrum and the technical support provided by the Barcelona Supercomputing Center (RES-IM-2025-2-0013).
All authors gratefully acknowledge the Universidad Politécnica de Madrid (www.upm.es) for providing computing resources on the Magerit Supercomputer.

\appendix

\section{Navier–Stokes Equations}
\label{sec:cNS}

The 3D Navier--Stokes equations including the Vreman model can be compactly written as:
\begin{equation}
\boldsymbol{u}_t+\nabla\cdot\ssvec{{F}}_e = \nabla\cdot\ssvec{F}_{v,turb},
\label{eq:compressibleNScompact}
\end{equation}
where $\boldsymbol{u}$ is the state vector of large-scale resolved conservative variables $\boldsymbol{u} = [ \rho , \rho v_1 , \rho v_2 , \rho v_3 , \rho e]^T$, $\ssvec{F}_e$ are the inviscid, or Euler fluxes,
\begin{equation}
\ssvec{F}_e = \left[\begin{array}{ccc} \rho v_1 & \rho v_2 & \rho v_3 \\
                                                                                \rho v_1^2 + p & \rho v_1v_2 & \rho v_1v_3 \\
                                                                                	\rho v_1v_2 & \rho v_2^2 + p & \rho v_2v_3 \\
                                                                                	\rho v_1v_3 & \rho v_2v_3 & \rho v_3^2 + p \\
                                                                                	\rho v_1 H & \rho v_2 H & \rho v_3 H
\end{array}\right],
\end{equation}
where $\rho$, $e$, $H=e+p/\rho$, and $p$ are the large-scale density, total energy, total enthalpy and pressure, respectively, and $\vec{v}=[v_1,v_2,v_3]^T$ is the large-scale resolved velocity components. Additionally, $\ssvec{F}_{v,turb}$ defines the viscous and turbulent fluxes,
\begin{equation}
\ssvec{F}_{v,turb}= \left[\begin{array}{ccc}0 & 0 & 0\\
\tau_{xx} & \tau_{xy} & \tau_{xz} \\
\tau_{yx} & \tau_{yy} & \tau_{yz} \\
\tau_{zx} & \tau_{zy} & \tau_{zz} \\
\sum_{j=1}^3 v_j\tau_{1j} + \kappa T_x& \sum_{j=1}^3 v_j\tau_{2j} + \kappa T_y& \sum_{j=1}^3 v_j\tau_{3j} + \kappa T_z
\end{array}\right],
\label{eq:viscousfluxes}
\end{equation}
where $\kappa$ is the thermal conductivity, $T_x, T_y$ and $T_z$ denote the temperature gradients and the stress tensor $\boldsymbol{\tau}$ is defined as $\boldsymbol{\tau} = (\mu+\mu_t)(\nabla \vec{v} + (\nabla \vec{v})^T) - 2/3(\mu+\mu_t) \boldsymbol{I}\nabla\cdot\vec{v}$, with $\mu$ the dynamic viscosity, $\mu_t$ the turbulent viscosity (in this work defined through the Vreman 
model).

\section{DGSEM}
\label{sec:DGSEM}

The physical domain is tessellated into non-overlapping curvilinear hexahedral elements, $e$, which are mapped to a reference element, $el$, using a polynomial transfinite mapping relating physical coordinates $\vec{x}$ and reference coordinates $\vec{\xi}$. The transformed Navier–Stokes equations read:

\begin{equation}
J \boldsymbol{u}_t + \nabla_\xi \cdot \cssvec{F}_e = \nabla_\xi \cdot \cssvec{F}_v,
\label{eq:compressibleNScompact_transformed}
\end{equation}

where $J$ is the Jacobian of the mapping, $\nabla_\xi$ denotes derivatives in the reference space, and $\cssvec{F}$ are the contravariant fluxes \cite{2009:Kopriva}.  

Multiplying \eqref{eq:compressibleNScompact_transformed} by a smooth test function $\phi_j$ ($0 \le j \le P$, with $P$ the polynomial degree) and integrating over an element yields the weak form:

\begin{equation}
\int_{el} J \boldsymbol{u}_t \phi_j + \int_{el} \nabla_\xi \cdot \cssvec{F}_e \phi_j = \int_{el} \nabla_\xi \cdot \cssvec{F}_v \phi_j.
\end{equation}

Integrating the inviscid flux term by parts separates surface and volume contributions:

\begin{equation}
\int_{el} J \boldsymbol{u}_t \phi_j + \int_{\partial el} \cssvec{F}_e \cdot \hat{\mathbf{n}} \phi_j - \int_{el} \cssvec{F}_e \cdot \nabla_\xi \phi_j = \int_{el} \nabla_\xi \cdot \cssvec{F}_v \phi_j,
\end{equation}

where $\hat{\mathbf{n}}$ is the outward unit normal on element faces. Discontinuous inviscid fluxes at element interfaces are replaced by a numerical Riemann flux, $\cssvec{F}_e^\star$, to ensure inter-element coupling:

\begin{equation}
\int_{el} J \boldsymbol{u}_t \phi_j + \int_{\partial el} \cssvec{F}_e^\star \cdot \hat{\mathbf{n}} \phi_j - \int_{el} \cssvec{F}_e \cdot \nabla_\xi \phi_j = \int_{el} \nabla_\xi \cdot \cssvec{F}_v \phi_j.
\end{equation}

Viscous terms can be integrated by parts as well, leading to formulations such as BR1, BR2, or interior penalty; here, we retain the simple volume form for clarity. The final DGSEM discretization approximates solutions and fluxes with polynomials of degree $P$ and evaluates volume and surface integrals using Gaussian quadrature. Gauss--Lobatto points enable energy/entropy-stable split forms, while Gauss–Legendre points provide higher quadrature accuracy.  

\section{Effect of the time-step size}
\label{sec:app_dt}

\textcolor{black}{To verify that the results of the $16P7$ configuration, mainly used throughout this work, are not affected by the temporal discretization, an additional simulation was performed using half the baseline time step. Specifically, the Roe--Chandrasekar iLES case with $\Delta t=2\times10^{-4}$ was repeated with $\Delta t=1\times10^{-4}$, while keeping all spatial discretization parameters unchanged.}

\begin{figure}[h!]
    \centering
    \begin{subfigure}[t]{0.49\textwidth}
        \includegraphics[width=\textwidth]{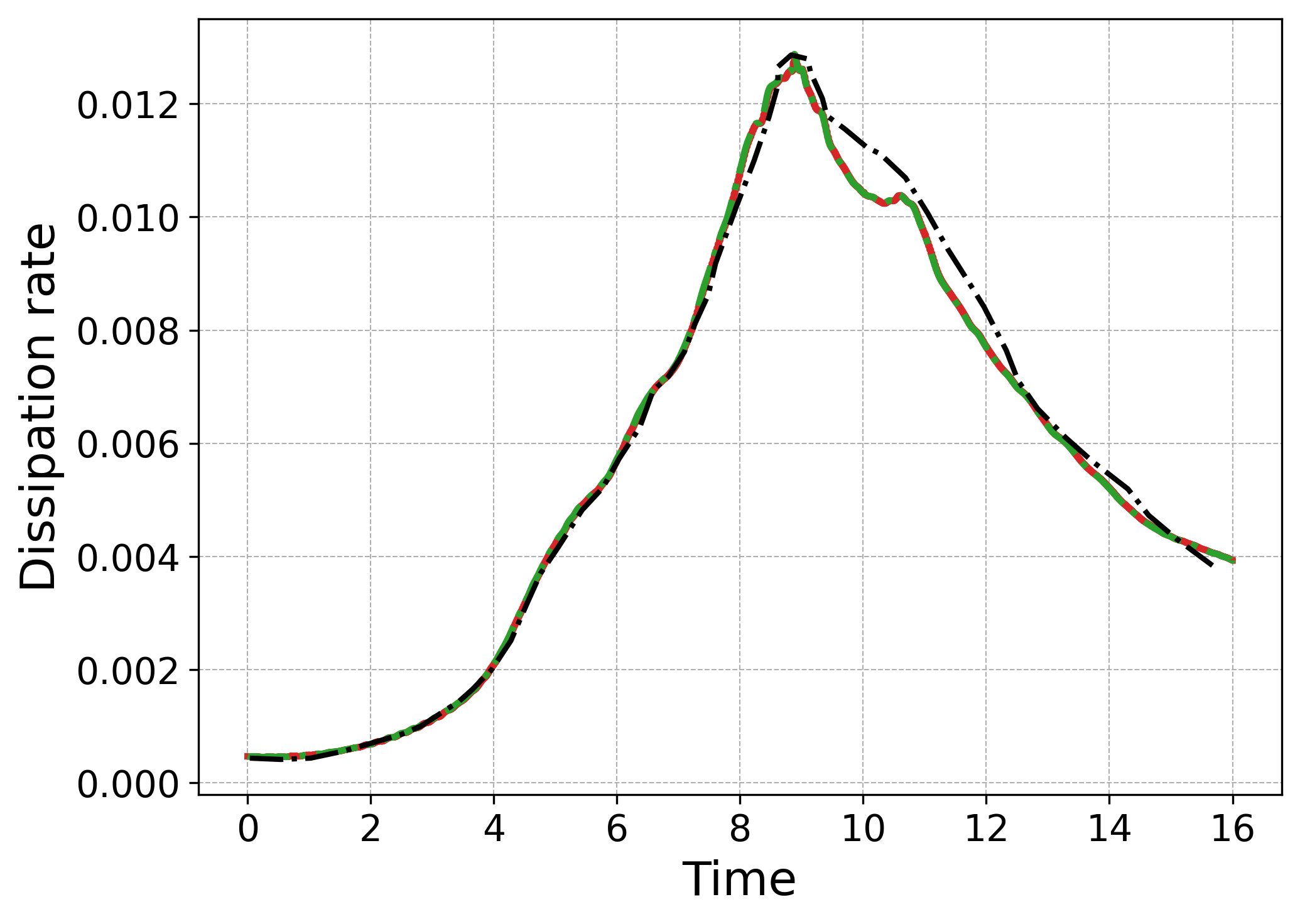}
        \caption{Kinetic energy dissipation rate.}
        \label{fig:appendix_dt_kinenrate}
    \end{subfigure}
    \hfill
    \begin{subfigure}[t]{0.49\textwidth}
        \includegraphics[width=\textwidth]{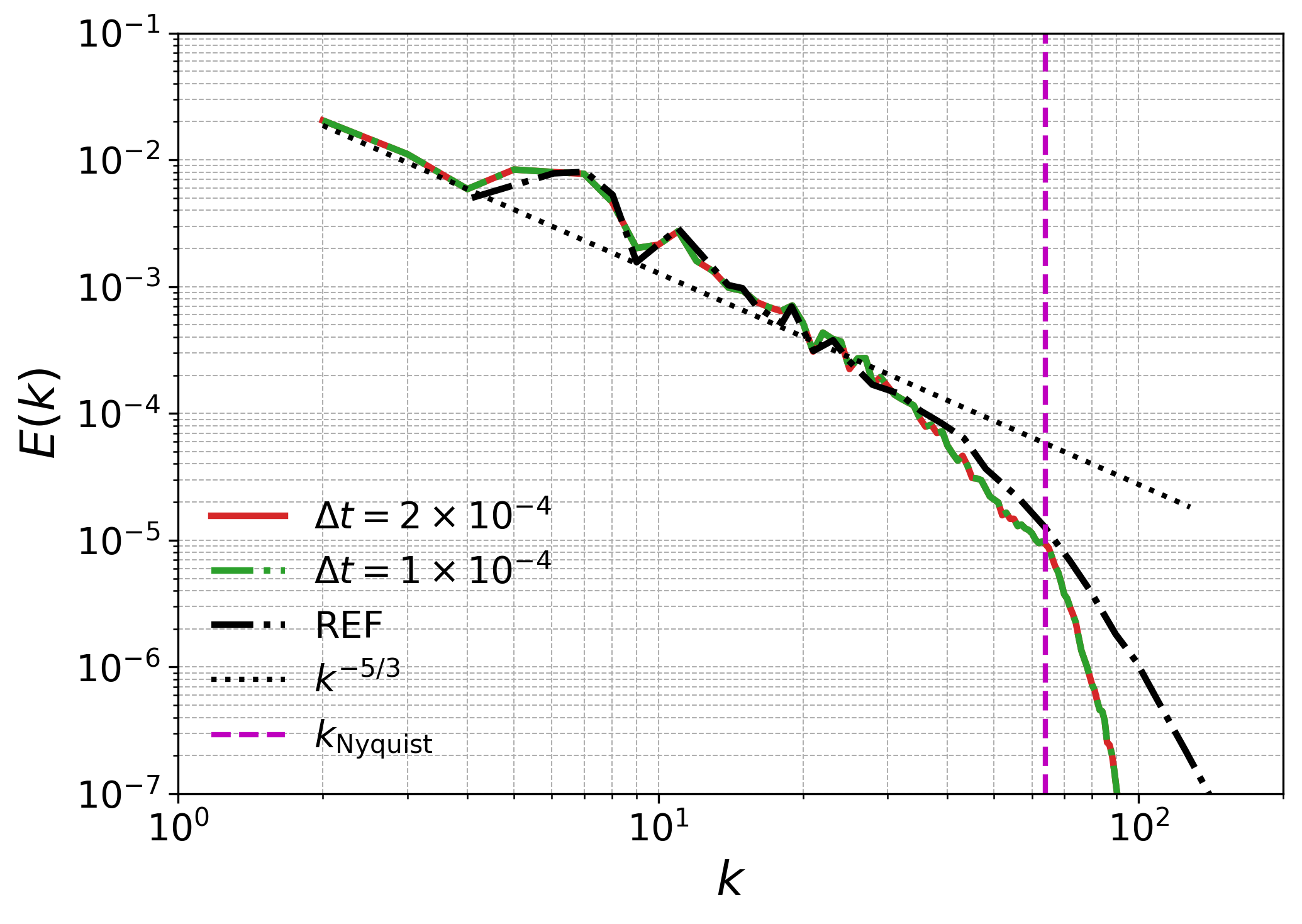}
        \caption{Kinetic energy spectrum at $t/t_c=9$.}
        \label{fig:appendix_dt_spectra}
    \end{subfigure}
    \caption{Time-step sensitivity analysis for the $16P7$ Roe--Chandrasekar iLES configuration. The baseline simulation with $\Delta t=2\times10^{-4}$ is compared with an additional simulation using $\Delta t=1\times10^{-4}$.}
    \label{fig:appendix_dt}
\end{figure}

\textcolor{black}{Figure~\ref{fig:appendix_dt} compares both simulations in terms of the kinetic energy dissipation rate and the kinetic energy spectrum at $t/t_c=9$. The dissipation histories are virtually indistinguishable, including the peak dissipation region. Likewise, the energy spectra show no relevant differences across the entire wavenumber range.}

\textcolor{black}{These results confirm that the baseline time step used for the $16P7$ simulations is sufficiently small, as is the typical situation when explicit time integration is used. Therefore, reducing the time step further does not improve the quantities analyzed in this work and only increases the computational cost.}

\section{Energy spectra computation}
\label{sec:app_fourier}
\textcolor{black}{
The energy spectra are computed by first evaluating the DG solution on an
auxiliary equispaced Cartesian grid. This post-processing step is needed because
the DG solution is represented by elementwise polynomials and is discontinuous
across element interfaces, while the Fourier Transform requires a single-valued field sampled
on a uniform periodic grid.}

\textcolor{black}{
At each point of the auxiliary grid, the element containing the point is
identified and the local DG polynomial is evaluated there. The velocity field is
then obtained from the conservative variables as
$(u,v,w)=(\rho u,\rho v,\rho w)/\rho$. No averaging between neighboring
elements and no additional filtering or smoothing are applied. 
We apply a multi-dimensional Fast Fourier Transform (FFT) to convert the velocity field from physical space into wavenumber (spectral) space. The isotropic
kinetic-energy spectrum is then obtained in the standard way by summing the absolute squares of the Fourier coefficients for each velocity component,
\[
\frac{1}{2}\left(|\hat{u}|^2+|\hat{v}|^2+|\hat{w}|^2\right)
\]
for each discrete wavenumber and then integrating over spherical shells (or omnidirectional averaging).
Thus, the reported spectra correspond to an FFT-based spectral analysis of the
elementwise DG polynomial solution sampled on an auxiliary uniform periodic grid.
}

\section*{Declaration of generative AI and AI-assisted technologies in the writing process}
During the preparation of this work the authors used OpenAI’s ChatGPT to correct grammar or spelling mistakes and to improve readability. After using this tool/service, the authors reviewed and edited the content as needed and take full responsibility for the content of the publication.

\end{document}